\documentclass[12pt]{article}
\usepackage{cite}
\usepackage{a4wide}
\usepackage{epsf}
\usepackage{epsfig}
\usepackage{color}

\def\nn{\nonumber}

\def\la{\langle}
\def\ra{\rangle}
\def\<{\langle}
\def\>{\rangle}

\def\l{\left}
\def\r{\right}

\def\Tr{\,{\rm Tr}\,}
\def\nn{\nonumber}

\def\beq{\begin{equation}}
\def\eeq{\end{equation}}
\def\bea{\begin{eqnarray}}
\def\eea{\end{eqnarray}}

\def\mev{\mbox{ MeV}}
\def\gev{\mbox{ GeV}}
\def\msbar{\overline{\mbox{MS}}}
\def\ln{{\mathrm{ln}}}
\def\op{{\mathcal{O}}}

\def\ampl{{\mathcal{M}}}

\def\vettp{\vec{p\,}}
\def\vetto{\vec{0}\,}
\def\ditre{$\Delta I=3/2\ $}

\begin{document}

\begin{titlepage}

\begin{flushright}\small{
DESY 04-213, FTUV-04-1106, IFIC/04-66,\\ INT-Pub 04-29,
LPT-Orsay/04-114, NT@UW-04-025,\\ ROMA-1392/04, RM3-TH/04-23, SHEP 0434, UW/PT
04-22}
\end{flushright}

\boldmath
\begin{center}
\Large\bf An exploratory lattice study of $\Delta I=3/2$
$K\rightarrow\pi\pi$ decays at next-to-leading order in the chiral
expansion
\end{center}
\unboldmath

\vspace{0.25cm}
\begin{center}
{\bf Philippe Boucaud}$^{a}$, {\bf Vicent Gim\'enez}$^{b}$, {\bf
C.-J. David Lin}$^{c}$, {\bf Vittorio Lubicz}$^{d}$, {\bf Guido
Martinelli}$^{e}$, {\bf Mauro Papinutto}$^{f}$, {\bf Chris T.
Sachrajda}$^{g}$
\\[0.8cm]
{\sl $^{a}$ Universit\'e de Paris Sud, LPT (B\^at. 210),
Centre d'Orsay, 91405 Orsay-Cedex, France}\\[0.37cm]
{\sl $^{b}$ Dep. de F\'isica Te\`orica and IFIC, Univ. de
Val\`encia, Dr. Moliner 50, E-46100, Burjassot, Val\`encia,
Spain}\\[0.37cm]
{\sl $^{c}$ Department of Physics, University of Washington,
Seattle, WA-98195-1550, USA \& Institute for Nuclear Theory,
University of Washington,
Seattle, WA-98195-1560, USA}\\[0.37cm]
{\sl $^{d}$ Dipartimento di Fisica, Universit\'a di Roma Tre and
INFN, Sezione di Roma III,\\Via della Vasca Navale 84, I-00146
Roma, Italy}\\[0.37cm]
{\sl $^{e}$ Dipartimento di Fisica, Universit\'a di Roma
``La Sapienza'' and\\INFN, Sezione di Roma, P.le A. Moro 2,
I-00185 Roma, Italy}\\[0.37cm]
{\sl $^{f}$ NIC/DESY Zeuthen, Platanenalle 6, D-15738 Zeuthen,
Germany}\\[0.7cm]
{\sl $^{g}$ School of Physics and Astronomy, University of Southampton,\\
Southampton SO17 1BJ, England}\\[0.37cm]
\end{center}

\begin{abstract}
\noindent We present the first direct evaluation of $\Delta I =
3/2$ $K\rightarrow\pi\pi$ matrix elements with the aim of
determining all the low-energy constants at NLO in the chiral
expansion. Our numerical investigation demonstrates that it is
indeed possible to determine the $K\to\pi\pi$ matrix elements
directly for the masses and momenta used in the simulation with
good precision. In this range however, we find that the matrix
elements do not satisfy the predictions of NLO chiral perturbation
theory. For the chiral extrapolation we therefore use a hybrid
procedure which combines the observed polynomial behaviour in
masses and momenta of our lattice results, with NLO chiral
perturbation theory at lower masses. In this way we find stable
results for the quenched matrix elements of the electroweak
penguin operators ($_{I=2}\la \pi\pi|\op_8|K^0\rangle= (0.68\pm
0.09)\gev^3 $ and $_{I=2}\la \pi\pi|\op_7|K^0\rangle= (0.12\pm
0.02)\gev^3$ in the NDR-$\overline{\textrm{MS}}$ scheme at the
scale 2\,GeV), but not for the matrix elements of ${\cal O}_4$
(for which there are too many Low-Energy Constants at NLO for a
reliable extrapolation). For all three operators we find that the
effect of including the NLO corrections is significant (typically
about 30\%). We present a detailed discussion of the status of the
prospects for the reduction of the systematic uncertainties.
\end{abstract}
\vfill\noindent

\end{titlepage}
\section{Introduction}
\label{sec:intro} The importance of a theoretical understanding of
$K\to\pi\pi$ decays is underlined by the recent measurements of a
non-zero value for the $\varepsilon^{\prime}/\varepsilon$
parameter \cite{Batley:2002gn,Alavi-Harati:2002ye} (the first
confirmed observation of direct CP violation) as well as the
long-standing puzzle of the $\Delta I = 1/2$ rule. Within the
Standard Model of particle physics, the major difficulty is to
quantify the non-perturbative QCD effects. Lattice simulations are
being used in attempts to overcome this problem, but the results
obtained in this way currently have large systematic errors. A
major source of uncertainty is induced by the chiral
extrapolation. Lattice simulations are performed with $u$- and
$d$-quark masses typically in the range $m_{s}/2 \leq m_{u,d} \leq
m_{s}$ (where $m_{s}$ is the strange-quark mass), and the physical
amplitudes are obtained by combining the lattice measurements with
Chiral Perturbation Theory ($\chi$PT). Previous calculations of
$K\to\pi\pi$ matrix elements were performed at leading-order (LO)
in the chiral expansion. Following some early attempts to evaluate
$K\to\pi\pi$ matrix elements directly~\cite{Gavela:1987bd,
Bernard:1988zj,Bernard:1989mv,bernard_tasi}, lattice estimates of
the $\Delta I=3/2$ $K\to\pi\pi$ matrix elements
$\la\pi^{+}\pi^{0}|\op^{3/2}_{7,8}| K^{+}\ra$~\cite{gupta_bk_97,
conti_bk_98, Lellouch:1998sg, Blum:2001xb, Noaki:2001un} have
generally been obtained by calculating the $K\to\pi$ matrix
elements $\la\pi^{+}|\op^{3/2}_{7,8}|K^{+}\ra$ and using the
soft-pion theorem $(-i)\la\pi|\op|K\ra/f_{\pi}\sim\la\pi\pi
|\op|K\ra$ \cite{bernard_chpt_85}. An exception to this has been
the computation of the $K\to\pi\pi$ matrix element of the operator
${\cal O}_+$ (defined in eq.\,(\ref{eq:o_plus_def}) below), with
all three mesons at rest~\cite{jlqcd_o4_98}.

We have previously proposed improving the precision of lattice
determinations of the decay amplitudes by computing directly
$K\to\pi\pi$ matrix elements at unphysical kinematics and
combining them with $\chi$PT at next-to-leading order
(NLO)~\cite{Boucaud:2001mg, Lin:2002nq}. By comparing the
behaviour of the computed matrix elements as functions of the
mesons' masses and momenta with the predictions of chiral
perturbation theory at NLO, we can, at least in principle,
determine the low-energy constants and hence evaluate the physical
amplitudes. In this paper we perform an exploratory numerical
study to investigate the extent to which such a programme can be
carried out in practice. In particular, we evaluate the matrix
elements $\la \pi^{+}\pi^{0}|\op_{i}|K^{+}\ra$ for the following
$\Delta I = 3/2$ operators $\op_{i}$:
\bea
\label{eq:o4_def}
 \op_{4} &=& (\bar{s}_{\alpha}d_{\alpha})_{L}
     \l [(\bar{u}_{\beta}u_{\beta})_{L} -
         (\bar{d}_{\beta}d_{\beta})_{L}\r ] +
       (\bar{s}_{\alpha}u_{\alpha})_{L}(\bar{u}_{\beta}d_{\beta})_{L}, \\
\label{eq:o7_32_def}
 \op^{3/2}_{7} &=& (\bar{s}_{\alpha}d_{\alpha})_{L}
     \l [(\bar{u}_{\beta}u_{\beta})_{R} -
         (\bar{d}_{\beta}d_{\beta})_{R}\r ] +
       (\bar{s}_{\alpha}u_{\alpha})_{L}(\bar{u}_{\beta}d_{\beta})_{R}, \\
\label{eq:o8_32_def}
 \op^{3/2}_{8} &=& (\bar{s}_{\alpha}d_{\beta})_{L}
     \l [(\bar{u}_{\beta}u_{\alpha})_{R} -
         (\bar{d}_{\beta}d_{\alpha})_{R}\r ] +
       (\bar{s}_{\alpha}u_{\beta})_{L}(\bar{u}_{\beta}d_{\alpha})_{R},
\eea
where $\alpha$ and $\beta$ are colour indices and
$(\bar{\psi}_{1}\psi_{2})_{L,R} =
\bar{\psi}_{1}\gamma_{\mu}(1\mp\gamma_{5})\psi_{2}$.

The matrix elements of the $\Delta I = 3/2$ operator $\op_{4}$,
first introduced in ref.\,\cite{Shifman:1977tn}, give dominant
contributions to CP-conserving $K\rightarrow\pi\pi$ decays.
$\op_{4}$ is proportional to the $\Delta I = 3/2$ component of the
operator
\beq
\label{eq:o_plus_def}
 \op_{+} = \frac{1}{2}\left [
  \left(\bar{s}_{\alpha}d_{\alpha}\right)_{L}
  \left(\bar{u}_{\alpha}u_{\alpha}\right)_{L}
  +
  \left(\bar{s}_{\alpha}u_{\alpha}\right)_{L}
  \left(\bar{u}_{\alpha}d_{\alpha}\right)_{L}
 \right ] ,
\eeq
whose matrix element, $\la \pi^{+}\pi^{0}|\op_{+}|K^{+}\ra = \la
\pi^{+}\pi^{0}|\op_{4}|K^{+}\ra /3$, is an important theoretical
input to the study of the $\Delta I = 1/2$ rule. A consequence of
chiral symmetry is that this matrix element can be related to $\la
K^{0}|(\bar{s}d)_{L}(\bar{s}d)_{L}|\bar{K}^{0}\ra$
\cite{donoghue_bk_82}, which contains the non-perturbative QCD
effects in $K^{0}{-}\bar{K}^{0}$ mixing (i.e. the two matrix
elements are given in terms of the same low-energy constants).

$\op^{3/2}_{7}$ and $\op^{3/2}_{8}$ are proportional to the
$\Delta I = 3/2$ component of the electroweak penguin operators,
\begin{equation}
\label{eq:o78_def}
 \op_{7} = \frac{3}{2}(\bar{s}_{\alpha}d_{\alpha})_{L}
       \sum_{q} e_{q} (\bar{q}_{\beta}
       q_{\beta})_{R}\qquad\textrm{and}\qquad
 \op_{8} = \frac{3}{2}(\bar{s}_{\alpha}d_{\beta})_{L}
    \sum_{q} e_{q} (\bar{q}_{\beta} q_{\alpha})_{R}\ .
\end{equation}
These operators, although suppressed by a factor of
$\alpha_{\mathrm{em}}/\alpha_{s}$ compared to QCD penguin
operators, can contribute significantly in the theoretical
prediction of $\varepsilon^{\prime}/\varepsilon$
\cite{bijnens_wise_ew_84, flynn_randall_ew_89}. Recent estimates
\cite{Bosch:1999wr, Ciuchini:1999xh, Bertolini:2000dy, Pallante:2001he, Buras:2003zz} suggest that $\op_{8}$
has a significant negative contribution to
$\varepsilon^{\prime}/\varepsilon$.

In principle, by using finite-volume techniques, it is possible to
calculate $K\rightarrow\pi\pi$ amplitudes directly from lattice
QCD \cite{Lellouch:2000pv,Lin:2001ek}, without using chiral
perturbation theory. However, in order to obtain the physical
amplitude from simulations with periodic boundary conditions for
the fields, one requires lattice sizes of order 6\,fm or larger,
which is numerically very demanding~\footnote{However, see refs.
\cite{Kim:2002np} and \cite{deDivitiis:2004rf} for the possibility
of reducing the required size of the volume by using different
boundary conditions.}. In addition for $\Delta I=1/2$ decays, the
lack of unitarity and enhanced long-distance effects in quenched
and partially quenched QCD suggest that one needs to use full QCD
to determine these matrix elements~\cite{Lin:2002aj,
Bernard:1996ez,Lin:2003tn}. This problem is much less severe for
the $\Delta I=3/2$ decays studied in this paper. For matrix
elements as complicated as those contributing to $K\to\pi\pi$
decay amplitudes, unquenched simulations on sufficiently large
volumes and on fine lattices are unlikely in the near future. It
will therefore continue to be necessary to combine $\chi$PT with
lattice simulations for the foreseeable future.

In this paper we present the first lattice study of the matrix
elements $\la\pi^{+}\pi^{0}|\op_{i}|K^{+}\ra$ ($\op_{i}=\op_{4},
\op^{3/2}_{7,8}$) at next-to-leading order (NLO) in the chiral
expansion.  We follow the strategy proposed in refs.
\cite{Boucaud:2001mg, Lin:2002nq}. Under the
$SU(3)_{\mathrm{L}}\times SU(3)_{\mathrm{R}}$ flavour
transformation, the operator $\op_{4}$ is in the
$(27_{\mathrm{L}},1_{\mathrm{R}})$ irreducible representation and
$\op_{7,8}$ (and hence $\op^{3/2}_{7,8}$) are in the
$(8_{\mathrm{L}},8_{\mathrm{R}})$ irreducible representation. This
means that the chiral expansion for the matrix elements of
$\op_{4}$ starts at $\op(p^{2})$, while that for $\op_{7,8}$ and
$\op^{3/2}_{7,8}$ starts at $\op(p^{0})$. In
sec.~\ref{sec:strategy} we recall the demonstration that by
allowing one of the pions to carry non-zero spatial momentum we
are able, in principle at least, to determine these matrix
elements at NLO in the chiral expansion~\cite{Boucaud:2001mg,
Lin:2002nq}.

Although we were able to determine the matrix elements
$\la\pi^{+}\pi^{0}|\op_{4}|K^{+}\ra$ at the quark masses at which
the simulations were performed, we were unable to perform the
chiral extrapolation with any degree of confidence. For this
matrix element there is one low-energy constant at lowest order
and six more at NLO and it was not possible to determine so many
parameters with sufficient precision by using our lattice data.
Nevertheless, our study already reveals some qualitative feature
of this matrix element, which will be discussed in detail in
sec.\,\ref{sec:analysis}. For
$\la\pi^{+}\pi^{0}|\op^{3/2}_{7,8}|K^{+}\ra$ there are fewer
low-energy constants and we were able to estimate the physical
matrix elements. Our best estimates for these matrix elements
renormalized in the NDR-$\msbar$ scheme at $2\gev$ are (see
sec.~\ref{sec:results}), \bea _{I=2}\la
\pi\pi|\op_8|K^0\rangle^{\textrm{\scriptsize{NLO}}}_{\textrm{\scriptsize{phys}}}&=&
0.68\,(6)\,(4)\,(5)\gev^3\nn\\
_{I=2}\la
\pi\pi|\op_7|K^0\rangle^{\textrm{\scriptsize{NLO}}}_{\textrm{\scriptsize{phys}}}&=&
0.12\,(1)\,(1)\,(1)\gev^3\,. \label{eq:best2}\eea The first error
is statistical, the second is due to the uncertainties in the
non-perturbative matching to the continuum renormalization scheme
and the third is due to the chiral extrapolations. In addition to
the errors quoted above, there are uncertainties due to quenching,
possible scaling violations and finite-volume effects which are
more difficult to estimate reliably. We discuss these in
sec.\,\ref{sec:systematics}.

The plan for the remainder of the paper is as follows. In the next
section we explain our strategy for the determination of all the
low-energy constants which are required for the evaluation of the
physical amplitudes at NLO in the chiral expansion. The parameters
and details of our simulation are presented in sec.~\ref{sec:sim}.
In sec.~\ref{sec:op_match} we explain our procedure for the
non-perturbative renormalization of the lattice operators. The
details of our analysis and a discussion of the systematic
uncertainties are presented in secs.~\ref{sec:analysis} and
\ref{sec:systematics} respectively. Section \ref{sec:results}
contains the main numerical results of this study and a comparison
with those from previous calculations.  Preliminary reports on
this work have been presented in refs. \cite{Boucaud:2001tx,
Becirevic:2002mm}. We present our conclusions in
sec.~\ref{sec:concs}. There is one appendix in which we study the
chiral behaviour of the $K\to\pi$ matrix elements obtained by the
CP-PACS collaboration~\cite{Noaki:2001un} and argue that different
choices of ansatz for the chiral extrapolation can lead to
significantly different results.

\section{Strategy for the Determination of the Physical
Amplitudes}\label{sec:strategy}

In this study we follow the strategy presented in refs.
\cite{Boucaud:2001mg, Lin:2002nq} for the determination of the
physical amplitudes from lattice simulations combined with
$\chi$PT at NLO. In order to be able to extract all the low-energy
constants at NLO, we must evaluate $K\to\pi\pi$ matrix elements
for a range of unphysical masses and momenta. We choose to use the
SPQR kinematics, in which one of the final-state pions has a
non-zero spatial momentum and the other two mesons are at rest.

The chiral expansion for the matrix elements
$\ampl^{(4)}_{\mathrm{SPQR}}\equiv\la\pi^{+}\pi^{0}|\op_{4}|K^{+}\ra$
and $\ampl^{(7,8)}_{\mathrm{SPQR}}\equiv
\la\pi^{+}\pi^{0}|\op^{3/2}_{7,8}|K^{+}\ra$ in the SPQR
kinematics, to NLO and in infinite
volume is:
\begin{eqnarray}
 \ampl^{(4)}_{\mathrm{SPQR}}
 & = &-\frac{6\sqrt{2}}
 {f_K\,f_\pi^2}\Bigg\{\alpha\Big(E_{\pi}M_{\pi}+
  \frac12M_K\,(E_{\pi} + M_{\pi})
  +[\textrm{chiral logs}]^{(4)}_{\mathrm{SPQR}}\Big)\nonumber\\
  &&\hspace{0in}+ 4 \beta_{2} M^{4}_{\pi}
  +(4\beta_{4}+2\beta_{7})E_{\pi}M^{3}_{\pi}
  +(\beta_{4}-\beta_{5}+\beta_{7})M^{3}_{\pi}M_{K}\nonumber\\
  &&\hspace{-0in}+ (\beta_{4}-\beta_{5}+\beta_{7}+2\beta_{22})
              E_{\pi}M^{2}_{\pi}M_{K}
  + (-4\beta_{2}+8\beta_{24})M^{2}_{\pi}M^{2}_{K}
+ (\beta_{4}+2\beta_{7})E_{\pi}M^{3}_{K}\nonumber\\
 && \hspace{-0in}+ (-2\beta_{5}+4\beta_{7}+4\beta_{22})
           E_{\pi}M_{\pi}M^{2}_{K}
  + (\beta_{4}+2\beta_{7})M_{\pi}M^{3}_{K}\nonumber\\
  &&\hspace{-0in} +(-16\beta_{24})E^{2}_{\pi}M^{2}_{\pi}
  + 2\beta_{22}E^{2}_{\pi} M_{\pi} M_{K}
  + 8\beta_{24}E^{2}_{\pi}M^{2}_{K}\,
     \Bigg\}\,\nonumber\\
  & = &-\frac{6\sqrt{2}}
 {f^{3}}\Bigg\{\nonumber\\
 &&\alpha\Big\{\Big [ E_{\pi}M_{\pi}+
  \frac12M_K\,(E_{\pi} + M_{\pi})\Big ]
  [\textrm{chiral logs}]_{f^{3}}(M_{\pi},M_{K})
  +[\textrm{chiral logs}]^{(4)}_{\mathrm{SPQR}}\Big\}
  \nonumber\\
  &&\hspace{0in}+ 4 \beta_{2} M^{4}_{\pi}
  +\left ( 4\beta_{4}+2\beta_{7}
   -\frac{24L_{4}}{f^{2}}-\frac{16L_{5}}{f^{2}}\right )
  E_{\pi}M^{3}_{\pi}\nonumber\\
  & & +\left (\beta_{4}-\beta_{5}+\beta_{7}
  -\frac{12L_{4}}{f^{2}}-\frac{8L_{5}}{f^{2}}\right )
   M^{3}_{\pi}M_{K}\nonumber\\
  && + \left (\beta_{4}-\beta_{5}+\beta_{7}+2\beta_{22}
   -\frac{12L_{4}}{f^{2}}-\frac{8L_{5}}{f^{2}}\right )
              E_{\pi}M^{2}_{\pi}M_{K}
  + (-4\beta_{2}+8\beta_{24})M^{2}_{\pi}M^{2}_{K}\nonumber\\
 &&
+ \left (\beta_{4}+2\beta_{7}
  -\frac{24L_{4}}{f^{2}}-\frac{4L_{5}}{f^{2}}\right )
  E_{\pi}M^{3}_{K}\nonumber\\
 && \hspace{-0in}
+\left (-2\beta_{5}+4\beta_{7}+4\beta_{22}
  -\frac{48L_{4}}{f^{2}}-\frac{8L_{5}}{f^{2}}\right )
           E_{\pi}M_{\pi}M^{2}_{K}\nonumber\\
 && + \left (\beta_{4}+2\beta_{7}
    -\frac{24L_{4}}{f^{2}}-\frac{4L_{5}}{f^{2}}\right )
    M_{\pi}M^{3}_{K}\nonumber\\
  &&\label{eq:chiralo4spqr}
  \hspace{-0in} +(-16\beta_{24})E^{2}_{\pi}M^{2}_{\pi}
  + 2\beta_{22}E^{2}_{\pi} M_{\pi} M_{K}
  + 8\beta_{24}E^{2}_{\pi}M^{2}_{K}\,
     \Bigg\}\, ,\\
 & &\nonumber\\
 \ampl^{(7,8)}_{\mathrm{SPQR}}
 & = &\frac{4\sqrt{2}}{f_K\,f_\pi^2}\,\Bigg\{\gamma^{(7,8)}\,(1 +
 [\textrm{chiral logs}]^{\mathrm{EWP}}_{\mathrm{SPQR}} ) +
 2\Big(\delta^{(7,8)}_4+\delta^{(7,8)}_5+2\delta^{(7,8)}_6\Big)\,M_K^2
 \nonumber\\
 &&+\hspace{-0in}\frac12\Big(\delta^{(7,8)}_1
  -\delta^{(7,8)}_2-2\delta^{(7,8)}_3\Big)(M_\pi+E_\pi)M_K\nonumber\\
 &&+2\Big(2(\delta^{(7,8)}_4+\delta^{(7,8)}_5)+\delta^{(7,8)}_6\Big)M_\pi^2
  -\Big(\delta^{(7,8)}_1+\delta^{(7,8)}_2\Big)M_\pi E_\pi \Bigg\}\nonumber\\
& &\nonumber\\
 & = &\frac{4\sqrt{2}}{f^{3}}\,\Bigg\{\gamma^{(7,8)}\,\Big (1 +
 [\textrm{chiral logs}]^{\mathrm{EWP}}_{\mathrm{SPQR}}
 +[\textrm{chiral logs}]_{f^{3}}(M_{\pi},M_{K})\Big ) \nonumber\\
 &&+
 2\Big(\delta^{(7,8)}_4+\delta^{(7,8)}_5+2\delta^{(7,8)}_6
 -\frac{24 L_{4}}{f^{2}}-\frac{4 L_{5}}{f^{2}}\Big)\,M_K^2
 \nonumber\\
 &&+\hspace{-0in}\frac12\Big(\delta^{(7,8)}_1
  -\delta^{(7,8)}_2-2\delta^{(7,8)}_3\Big)(M_\pi+E_\pi)M_K
 \nonumber\\
 &&+2\Big(2(\delta^{(7,8)}_4+\delta^{(7,8)}_5)+\delta^{(7,8)}_6
 -\frac{12 L_{4}}{f^{2}}-\frac{8 L_{5}}{f^{2}}\Big)M_\pi^2
  -\Big(\delta^{(7,8)}_1+\delta^{(7,8)}_2\Big)M_\pi E_\pi \Bigg\}
  \,,
\label{eq:chiralo78spqr}
\end{eqnarray}
where $[\textrm{chiral logs}]^{(4)}_{\mathrm{SPQR}}$ and
$[\textrm{chiral logs}]^{\mathrm{EWP}}_{\mathrm{SPQR}}$ are
calculated in \cite{Lin:2002nq} in both full QCD and quenched QCD
(qQCD). With the exception of $f$, which is the pseudoscalar decay
constant in the chiral limit~\footnote{We work in the convention
for which the physical value of $f_\pi$ is 132\,MeV. For the
physical pion and kaon mass we have used the values $M_\pi=137$
MeV and $M_K=495$ MeV.}, the quantities appearing in
eqs.\,(\ref{eq:chiralo4spqr}) and (\ref{eq:chiralo78spqr}) all
correspond to the masses and energies used in the simulations
($M_\pi$ is the pion mass, $M_K$ the kaon mass and $E_\pi$ the
energy of the pion with non-zero spatial momentum). All
dimensionful quantities are taken to be in lattice units.
$\alpha$, $\beta_{i}$, $\gamma^{(7,8)}$ and $\delta^{(7,8)}_{i}$
are the unknown low-energy constants (LECs) in the chiral
expansion for these weak amplitudes. $L_{4,5}$ are the standard
Gasser-Leutwyler constants in the strong chiral Lagrangian
\cite{Gasser:1985gg}. $[\textrm{chiral logs}]_{f^{3}}$ represents
the chiral logarithms appearing in the renormalization of the
factor $1/f_Kf_\pi^2$ and are known for both QCD and
qQCD~\cite{Gasser:1985gg,Golterman:1997wb}.

The chiral expansion for the physical matrix elements,
$\ampl^{(4,7,8)}_{\mathrm{phys}}$ (that is
$\la\pi^{+}\pi^{0}|\op_{4}|K^{+}\ra$ and
$\la\pi^{+}\pi^{0}|\op^{3/2}_{7,8}|K^{+}\ra$ with the four-momenta
of the initial and final states equal), to NLO in infinite volume
is:
\bea \ampl^{(4)}_{\mathrm{phys}}&=&
  -\frac{6\sqrt{2}}
 {f_K\,f_\pi^2}\Bigg\{\alpha
 \bigg(\,M^{2}_K- M^{2}_{\pi}+[\textrm{chiral logs}]^{(4)}_{\mathrm{phys}}
 \bigg)\nonumber\\
 &&\hspace{-0in}+ (\beta_{4}-\beta_{5}+4\beta_{7}+ 2 \beta_{22})
     M^{4}_{K}\nonumber\\
  && + (4\beta_{2}-4\beta_{4}-2\beta_{7}-16\beta_{24})
     M^{4}_{\pi}\nonumber\\
 &&\hspace{-0in} +(-4\beta_{2}+3\beta_{4}+\beta_{5}-2\beta_{7}-2\beta_{22}
      +16\beta_{24}) M^{2}_{K}M^{2}_{\pi}\Bigg\}\ \nonumber\\
&=&
 -\frac{6\sqrt{2}}
 {f^{3}}\Bigg\{\alpha
 \bigg(\,(M^{2}_K- M^{2}_{\pi})
   [\textrm{chiral logs}]_{f^{3}}(M_{\pi},M_{K})
  +[\textrm{chiral logs}]^{(4)}_{\mathrm{phys}}
 \bigg)\nonumber\\
 &&\hspace{-0in}+ \left (\beta_{4}-\beta_{5}+4\beta_{7}+ 2 \beta_{22}
  - \frac{48 L_{4}}{f^{2}} - \frac{8 L_{5}}{f^{2}}\right )
     M^{4}_{K}\nonumber\\
  && + \left (4\beta_{2}-4\beta_{4}-2\beta_{7}-16\beta_{24}
     + \frac{24 L_{4}}{f^{2}} + \frac{16 L_{5}}{f^{2}}\right )
     M^{4}_{\pi}\nonumber\\
 &&\hspace{-0in} +\left (-4\beta_{2}+3\beta_{4}+\beta_{5}-2\beta_{7}-2\beta_{22}
      +16\beta_{24}
      + \frac{24 L_{4}}{f^{2}} - \frac{8 L_{5}}{f^{2}}\right )
    M^{2}_{K}M^{2}_{\pi}\Bigg\}\ ,
\label{eq:chiralo4phys}\\
& &\nonumber\\
\ampl^{(7,8)}_{\mathrm{phys}}&=&
\frac{4\sqrt{2}}{f_K\,f_\pi^2}\,\Bigg\{\gamma^{(7,8)}\,(1 +
 [\textrm{chiral logs}]^{\mathrm{EWP}}_{\mathrm{phys}}) \nonumber\\
 &&+\hspace{-0in}\Big(
 -\delta^{(7,8)}_2-\delta^{(7,8)}_3+2(\delta^{(7,8)}_4
 +\delta^{(7,8)}_5)+4\delta^{(7,8)}_6\Big)\,M_K^2\nonumber\\
&&+\Big(
 \delta^{(7,8)}_1+\delta^{(7,8)}_2+4(\delta^{(7,8)}_4
 +\delta^{(7,8)}_5)+2\delta^{(7,8)}_6\Big)M_\pi^2\Bigg\}
\nonumber\\
&=& \frac{4\sqrt{2}}{f^3}\,\Bigg\{\gamma^{(7,8)}\,\Big (1 +
 [\textrm{chiral logs}]^{\mathrm{EWP}}_{\mathrm{phys}}
 +[\textrm{chiral logs}]_{f^{3}}(M_{\pi},M_{K})\Big ) \nonumber\\
 &&+\hspace{-0in}\Big(
 -\delta^{(7,8)}_2-\delta^{(7,8)}_3+2(\delta^{(7,8)}_4
 +\delta^{(7,8)}_5)+4\delta^{(7,8)}_6
 -\frac{48 L_{4}}{f^{2}}-\frac{8 L_{5}}{f^{2}}\Big)\,M_K^2\nonumber\\
&&+\Big(
 \delta^{(7,8)}_1+\delta^{(7,8)}_2+4(\delta^{(7,8)}_4
 +\delta^{(7,8)}_5)+2\delta^{(7,8)}_6
 -\frac{24 L_{4}}{f^{2}}-\frac{16 L_{5}}{f^{2}}
\Big)M_\pi^2\Bigg\}\,. \label{eq:chiralo78phys} \eea
We use the formulae for $[\textrm{chiral
logs}]^{(4)}_{\mathrm{phys}}$ and $[\textrm{chiral
logs}]^{\mathrm{EWP}}_{\mathrm{phys}}$ for both full QCD and
quenched QCD (qQCD) presented in \cite{Lin:2002nq}.

\bigskip

In eqs. (\ref{eq:chiralo4spqr})\,--\,(\ref{eq:chiralo78phys}), we
have written the amplitudes using two expressions which are
equivalent at NLO in the chiral expansion. The difference
corresponds to the two ways that we incorporate the chiral
corrections to $1/f^{3}$ at NLO. This leads to two procedures for
obtaining the LECs by fitting $\ampl^{(4,7,8)}_{\mathrm{SPQR}}$ to
a function of the masses and energies:
\begin{enumerate}
\item we multiply the amplitudes given by the first expressions in
each of eqs.~(\ref{eq:chiralo4spqr}) and (\ref{eq:chiralo78spqr})
by $f_Kf_\pi^2$ computed at the same values of the quark masses,
and obtain the parameters $\alpha$, the $\beta$'s, $\gamma$ and
the $\delta$'s by fitting; \item we determine the ratios
$\alpha/f^3$, the $\beta/f^3$'s, $\gamma/f^3$ and the
$\delta/f^3$'s by fitting the second expressions in
eqs.~(\ref{eq:chiralo4spqr}) and (\ref{eq:chiralo78spqr}).
\end{enumerate}
Notice that in the second approach, we do not need to determine
the Gasser-Leutwyler constants $L_{4}$ and $L_{5}$ in order to
obtain the matrix elements, because these constants can be
absorbed into the LECs $\beta_{i}$, and $\delta^{(7,8)}_{i}$ via
the re-definition:
\bea
\label{eq:redef}
 \beta_{4} &\rightarrow& \beta_{4}^{\prime} = \beta_{4} +
  \frac{4 L_{5}}{f^{2}} ,\nonumber\\
 \beta_{5} &\rightarrow& \beta_{5}^{\prime} = \beta_{5} -
  \frac{4 L_{5}}{f^{2}} ,\nonumber\\
 \beta_{7} &\rightarrow& \beta_{7}^{\prime} = \beta_{7} +
  \frac{12 L_{4}}{f^{2}} ,\nonumber\\
 (\delta^{(7,8)}_{4} +\delta^{(7,8)}_{5}) &\rightarrow&
  (\delta^{(7,8)}_{4} + \delta^{(7,8)}_{5})^{\prime} =
  (\delta^{(7,8)}_{4} +\delta^{(7,8)}_{5}) + \frac{4L_{5}}{f^{2}} ,\nonumber\\
 \delta^{(7,8)}_{6} &\rightarrow&
  \delta^{(7,8)\prime}_{6} =
  \delta^{(7,8)}_{6} + \frac{12L_{4}}{f^{2}} .
\eea
The difference between these two procedures is at next-to-next-to
leading order, hence beyond the precision we are trying to reach
in this work. We find however, that performing the fits using the
second procedure leads to more stable fits and smaller statistical
errors. We therefore quote the results obtained using the second
procedure.
\section{Details of the Simulation}
\label{sec:sim}

The results presented in this paper were obtained from two sets of
data. Set 1 with 340 quenched gauge configurations and set 2 with
480 configurations, both on a $24^{3}\times64$ lattice at
$\beta=6.0$ (which corresponds to an inverse lattice spacing
$a^{-1}=1.98(6)\gev$ as determined from the masses of the $K$ and
$K^*$ mesons) but on which different valence quark masses have
been simulated. The gauge action is the standard Wilson plaquette
action, and the fermions are described by the
Sheikholeslami-Wohlert (SW) action \cite{sw_action} with the
coefficient $c_{\mathrm{SW}}$ tuned so that the action does not
contain $\op(a)$ artefacts to all orders in $\alpha_{s}$
\cite{NP_clover_action} (at our value of $\beta$,
$c_{\mathrm{SW}}=1.769$). The $\op(a)$ improvement of the lattice
four-quark operators requires mixing with operators of higher
dimension and is beyond the scope of this paper.

In tab.~\ref{tab:ps_mass} we list the masses of the pseudoscalar
mesons used in our calculations. For the \textit{pions}, for each
data set we take the three combinations with a degenerate
quark-antiquark pair. For the \textit{kaons} we take all
quark-antiquark combinations. For each pion the corresponding kaon
has one quark or antiquark with the same mass (the kaons
corresponding to each pion are listed in tab.~\ref{tab:ps_mass}).
Since we wish to exploit the strategy proposed in
refs.~\cite{Boucaud:2001mg, Lin:2002nq}, the kaon and one of the
final-state pions are at rest, while the spatial momentum of the
other pion is either $\vettp=\vec{0}$ or
$\vec{p}=\vettp_{\textrm{\scriptsize{min}}}\equiv(2\pi/L, 0, 0)$
(all physical quantities are averaged over the three directions).

\begin{table}
\begin{center}
\begin{tabular}{c|c|c|c|c|c}
\hline
\hline
$\#$&$\kappa_{1}$ & $\kappa_{2}$ & $aM_{P}$ &
$aE_{\pi}(\vettp_{\textrm{\scriptsize{min}}})$& meson \\
\hline \hline
\multicolumn{6}{c}{Set 1 (340 configurations)}\\
\hline
\hline
1a&0.13449 &0.13449&  0.2571(16)  &   0.3644(25) &  pion, kaon\\
2a&0.13449 &0.13250&  0.3902(13)  &   --& kaon \\
3a&0.13449 &0.12997&  0.5177(12) &   --& kaon \\
\hline
1b&0.13376 &0.13376&  0.3593(13)  &   0.4431(16) &  pion, kaon\\
2b&0.13376 &0.13146&  0.4823(10)  &   --& kaon \\
3b&0.13376 &0.12565&  0.7215(10) &   --& kaon \\
\hline
1c&0.13300 &0.13300&  0.4432(9)  &   0.5127(11) &  pion, kaon\\
2c&0.13300 &0.12994&  0.5853(8)  &   --& kaon \\
3c&0.13300 &0.12612&  0.7343(8) &   --& kaon \\
\hline
\hline
\multicolumn{6}{c}{Set 2 (480 configurations)}\\
\hline
\hline
1a&0.1344 & 0.1344 & 0.2716(15)  & 0.3770(23)& pion, kaon\\
2a&0.1344 & 0.1337 & 0.3224(14) & --& kaon \\
3a&0.1344 & 0.1328 & 0.3785(12) & --& kaon \\
4a&0.1344 & 0.1318 & 0.4338(11) & --& kaon \\
5a&0.1344 & 0.1308 & 0.4839(11) & --& kaon \\
6a&0.1344 & 0.1299 & 0.5259(11) & --& kaon \\
7a&0.1344 & 0.1282 & 0.5995(11) & --& kaon \\
\hline
1b&0.1337 & 0.1337 & 0.3663(12) & 0.4488(15)& pion, kaon \\
2b&0.1337 & 0.1344 & 0.3224(14) & --& kaon \\
3b&0.1337 & 0.1328 & 0.4173(10) & --& kaon \\
4b&0.1337 & 0.1316 & 0.4790(9) & --& kaon \\
5b&0.1337 & 0.1306 & 0.5260(8) & --& kaon \\
6b&0.1337 & 0.1295 & 0.5746(8) & --& kaon \\
\hline
1c&0.1328 & 0.1328 & 0.4642(8) & 0.5309(10)& pion, kaon \\
2c&0.1328 & 0.1344 & 0.3785(12) & --& kaon \\
3c&0.1328 & 0.1337 & 0.4173(10) & --& kaon \\
4c&0.1328 & 0.1316 & 0.5220(7) & --& kaon \\
5c&0.1328 & 0.1307 & 0.5625(7) & --& kaon \\
\hline
\hline
\end{tabular}
\end{center}
\caption{\label{tab:ps_mass}The $\kappa$ values and the
corresponding non-degenerate pseudoscalar masses $M_{P}$ for the
two sets of data. The {\it pions} correspond to the degenerate
cases 1a, 1b, 1c while the {\it kaons} respectively to
$\{1a,2a,\ldots\}$, $\{1b,2b,\ldots\}$, $\{1c,2c,\ldots\}$.
$E_{\pi}(\vettp_{\textrm{\scriptsize{min}}})$ is the energy
associated with a {\it pion} when it carries the spatial momentum
$\vettp_{\textrm{\scriptsize{min}}}$ of one lattice unit. All the
errors are statistical and obtained with the jacknife procedure.}
\end{table}

In order to determine the matrix elements of the lattice operators
$\bar\op_{i}=\bar\op_{4}, \bar\op^{3/2}_{7}$ and
$\bar\op^{3/2}_{8}$ (where the bar denotes the bare lattice
operator) we evaluate the correlation functions
\bea \label{eq:4q_correlator}
 C^{(i)}_{0}(t,\vec{p}) &=& \la 0| P_{\pi^{+}}(t_{\textrm{\scriptsize{fix}}},\vec{0})
                   P_{\pi^{0}}(t,\vec{p})\bar{\op}_{i}(0)
                   P^{\dagger}_{K^{+}}(-t_{\textrm{\scriptsize{fix}}},\vec{0})|0\ra\ ,
                   \nonumber\\
 C^{(i)}_{+}(t,\vec{p}) &=& \la 0| P_{\pi^{0}}(t_{\textrm{\scriptsize{fix}}},\vec{0})
                   P_{\pi^{+}}(t,\vec{p})\bar{\op}_{i}(0)
                   P^{\dagger}_{K^{+}}(-t_{\textrm{\scriptsize{fix}}},\vec{0})|0\ra\ ,
\eea
where
\beq
\label{eq:interpol_op}
 P_{P}(t,\vec{p})\equiv  \int d^{\,3}x\mbox{ }{\mathrm{e}}^{-i \vec{p}\cdot\vec{x}}
    \phi_{P}(\vec{x},t) ,
\eeq
and $\phi_{P}(\vec{x},t)$ is the local interpolating operator for
the pseudoscalar meson $P$ (e.g. $\phi_{\pi^+}(x)=
\bar{d}(x)\gamma^5u(x)$). In the correlation functions in
eq.~(\ref{eq:4q_correlator}), the spatial momentum $\vec{p}$ is
either zero or $\vettp_{\textrm{\scriptsize{min}}}$, while
$t_{\textrm{\scriptsize{fix}}}$ is the time for which the two
pions propagate and in our study it is fixed to be $12$. In a
preliminary study performed on 140 gauge configurations with pion
masses covering the range of the present simulation we considered
the three cases $t_{\textrm{\scriptsize{fix}}}=10$,
$t_{\textrm{\scriptsize{fix}}}=12$ and
$t_{\textrm{\scriptsize{fix}}}=14$. Observing no discrepancy we
concluded that $t_{\textrm{\scriptsize{fix}}}=12$ is an
appropriate choice.

In addition to the four-point functions in
eq.\,(\ref{eq:4q_correlator}), in order to eliminate the matrix
elements of the interpolating operators $P_{\pi^{+,0}}$ and
$P_{K^+}$ and to compute the pseudoscalar decay constant we need
to evaluate the two-point correlation functions
\bea
\label{eq:2pts_pp}
 C_{PP}(t,\vec{p}) &=& \la 0|P_{P}(t,\vec{p})\phi_P^{\dagger}(\vec{0},0)|0\ra,
\\
\label{eq:2pts_pa}
 C_{PA}(t,\vec{p}) &=& \la 0|P_{P}(t,\vec{p})A^{\dagger}_{0}(\vec{0},0)|0\ra,
\eea
where $A_{0}$ is the time component of the local axial-vector
current.

\bigskip

\section{Renormalization of the Operators}
\label{sec:op_match} To obtain the physical amplitudes,
lattice-regularised  operators have to be matched to some
continuum renormalization scheme (normally the $\msbar$ scheme) in
which the Wilson coefficients are calculated. If chiral symmetry
is preserved by the regularization, $\op^{3/2}_{7}$ and
$\op^{3/2}_{8}$ mix between themselves, while $\op_{4}$
renormalizes multiplicatively. In a Wilson-like regularization,
such as the one  adopted in this work, chiral symmetry is
explicitly broken and  the bare lattice operators
$\bar{\op}_{4}(a)$, $\bar{\op}^{3/2}_{7}(a)$ and
$\bar{\op}^{3/2}_{8}(a)$ mix with other dimension-6 operators with
different chiral properties~\footnote{In the $SU(2)$ isospin
limit, the operators do not mix with those of lower dimension, as
this mixing can only occur through penguin contractions,
corresponding to $\Delta I = 1/2$ transitions.}. Using $\cal CPS$
symmetry it can be shown that mixing with operators of different
chirality only occurs in the parity-even
sector~\cite{martinelli_4q_84,Draper,bernard_cps_lat87}. Therefore
the renormalization properties of the parity-odd $\Delta I=3/2$
operators considered here are not affected by the breaking of
chiral symmetry due to the lattice fermion action. The bare and
renormalized operators are therefore related by
\beq\label{eq:o4ren}
 \op_{4}(\mu) = {\cal Z}_{+}(\mu a) \bar{\op}_{4}(a),
\eeq
and
\bea
 \l ( \begin{array}{cc} \op^{3/2}_{7}(\mu)\\
    \op^{3/2}_{8}(\mu) \end{array} \r ) &=&
 \hat{{\cal Z}}(\mu a)
 \l ( \begin{array}{cc} \bar{\op}^{3/2}_{7}(a)\\
                        \bar{\op}^{3/2}_{8}(a) \end{array} \r )\nonumber \\
 &=& \l ( \begin{array}{cc} {\cal Z}_{77}(\mu a) & {\cal Z}_{78}(\mu a)\\
                            {\cal Z}_{87}(\mu a) & {\cal Z}_{88}(\mu a)
          \end{array} \r )
     \l ( \begin{array}{cc} \bar{\op}^{3/2}_{7}(a)\\
                            \bar{\op}^{3/2}_{8}(a) \end{array} \r ),
\label{eq:o78ren}\eea
where $\mu$ is the renormalization scale. The renormalization
constants depend, of course, on the continuum scheme used to
define the renormalized operators. In the following we will
consider the RI-MOM and the $\msbar$ schemes for which the Wilson
coefficients at the next-to-leading order (NLO) have been computed
in refs.~\cite{bur2,buras_ds1_93_1,buras_ds1_93_2,
Buras:1993dy,ciuchini_ds1_94,ciuchini_ds1_95,Ciuchini:1998bw}.

In perturbation theory at NLO, matching of lattice and
renormalized operators requires a one-loop lattice (and continuum)
calculation of the matrix elements of four-quark operators between
quark states. However, lattice perturbation theory frequently has
large coefficients and we know from many examples where a
non-perturbative determination of the renormalization constants is
possible that one-loop lattice perturbation theory may be
insufficiently precise. Moreover, it has been shown in
refs.~\cite{conti_bk_98,Lellouch:1998sg} that the large anomalous
dimension and the scheme-dependent constants in $\hat{{\cal Z}}$
result in significant uncertainties in the determination of the
matrix elements of $\op^{3/2}_{7,8}$~\footnote{The $\msbar$
two-loop anomalous dimensions are also large for the electroweak
penguins. One can also determine the lattice two-loop anomalous
dimensions and they are even larger than their $\msbar$
counterparts.}.
\par Several methods have been proposed in attempts to improve the accuracy
in the determination of the mixing coefficients. One of these
consists in evaluating the renormalization constants computed in
perturbation theory using an effective coupling which resums
tadpole contributions~\cite{parisi,lepage_mackenzie_ti_93} and
which is expected therefore to reduce higher order corrections; we
refer to this as Boosted Perturbation Theory (BPT) (there are also
more elaborate variations of this approach which we will not
consider here). However, as will be seen in
sec.~\ref{subsec:pertmatch} below, since the one-loop coefficients
in $\hat {\cal Z}$, and particularly in ${\cal Z}_{77}$ and ${\cal
Z}_{88}$, are so large, it is difficult to be confident that
higher order terms would not change the results significantly. It
is therefore important to determine the renormalization constants
non-perturbatively. We implement such a non-perturbative method,
in which these constants are determined from the computation of
Green functions between quark and gluon external states, as
proposed in ref.~\cite{rome_np_renorm}. This method has already
been extensively applied to the parity-even component of the
$\Delta S=2$ and $\Delta I=3/2$ operators. A detailed discussion
can be found in ref.~\cite{donini_4q_np_renorm}. In the
calculation of the physical amplitudes we will use both the
perturbative and the non-perturbative renormalization constants,
although our final results will be those obtained with the
non-perturbative method.

\subsection{ Perturbative matching}
\label{subsec:pertmatch}

The renormalization constants ${\cal Z}_{+}$ and $\hat{{\cal Z}}$
defined in eqs.\,(\ref{eq:o4ren}) and (\ref{eq:o78ren}) can be
related to the one-loop perturbative corrections of local quark
bilinears~\cite{martinelli_4q_84}. The general formulae relating
the renormalization constants of the bilinear operators to those
of the four-fermion operators in a generic lattice regularization
have been derived  in \cite{gupta_bk_97}. In this subsection we
determine the one-loop matching coefficients of the four-quark
operators $\op_4$ and $\op_{7,8}^{3/2}$ between the lattice
regularization which we are using and the $\msbar$-NDR,
$\msbar$-HV ('t Hooft-Veltman) and RI-MOM renormalization schemes.
We use the matching coefficients for the quark bilinears from
refs.~\cite{Gabrielli:1991us,german_2q_lat96} and combine them
with the formulae which relate the perturbative corrections to
four-quark and bilinear operators  and which were obtained using
Fierz identities and charge conjugation symmetry
\cite{gupta_bk_97,martinelli_4q_84}~\footnote{We take
$c_{\mathrm{SW}}=1$ in the lattice perturbation theory, which is
the consistent procedure at one-loop
level~\cite{german_2q_lat96}.}. For the RI-MOM scheme we find:
\bea\label{eq:zplusri}
 {\cal Z}^{RI-MOM}_{+} &=& 1+ \frac{\alpha_{s}}{4\pi}\l (-4\ln (\mu a)
 -37.517\r ),\\
 & & \nonumber \\
\label{eq:z78ri}
 \hat{{\cal Z}}^{RI-MOM} &=& \l ( \begin{array}{cc}
        1+\frac{\alpha_{s}}{4\pi}\l ( -2\,\ln(\mu a) - 34.604\r ) &
          \frac{\alpha_{s}}{4\pi}\l ( 6\,\ln(\mu a) -12.672\r )\\
          \frac{\alpha_{s}}{4\pi}\l (-2.386 \r) &
        1+\frac{\alpha_{s}}{4\pi}\l ( 16\,\ln(\mu a) -65.460\r )\\
        \end{array}\r ) \, .
\eea
In order to convert this result into the $\msbar$-NDR scheme or into the
$\msbar$-HV scheme we give the matching coefficients
$r_+^{NDR,HV}$ and $\hat r^{NDR,HV}$
\cite{ciuchini_ds1_95} defined through
\bea
\label{eq:matchRIMS1}
{\cal O}_4^{NDR,HV}&=&\l(1-\frac{\alpha_s(\mu)}{4
  \pi}\,r_+^{NDR,HV}\r){\cal O}_4^{RI-MOM},\\ \label{eq:matchRIMS11} ({\cal O}^{3/2}_i)^{NDR,HV}&=&\l(\delta_{ij}-\frac{\alpha_s(\mu)}{4
  \pi}\,\hat r_{ij}^{NDR,HV}\r)({\cal O}^{3/2}_j)^{RI-MOM}.
\eea
The values of $r_+^{NDR,HV}$ and $\hat r^{NDR,HV}$
are
\bea
\label{eq:matchRIMS2}
r_+^{NDR}=\frac{14}{3} - 8\, \ln\, 2\quad\quad \hat r^{NDR}=
 \l ( \begin{array}{cc} \frac{2}{3} + \frac{2}{3}\, \ln\, 2&\quad -2 - 2\, \ln\, 2\\
 2 -2\, \ln\, 2  &\quad -\frac{34}{3}+ \frac{2}{3}\, \ln\, 2\\ \end{array}\r )
\eea
and
\bea
\label{eq:matchRIMS3}
r_+^{HV}=-2 - 8\, \ln\, 2\quad\quad \hat r^{HV}=
 \l ( \begin{array}{cc} -\frac{8}{3} + \frac{2}{3}\, \ln\, 2&\quad -8 - 2\, \ln\, 2\\
 -2 -2\, \ln\, 2  &\quad -\frac{62}{3}+ \frac{2}{3}\, \ln\, 2\\ \end{array}\r
 )\,.
\eea
\par The coefficients in eqs.~(\ref{eq:zplusri})--(\ref{eq:z78ri}) are
very large. In estimating the numerical values of the
renormalization constants, we try to minimize the higher-order
contributions by using a boosted coupling constant defined by:
\beq \bar \alpha_{s}= \frac{\alpha_{s}^{(0)} }{\cal P}\,
,\hspace{0.2in} \textrm{where}\hspace{0.2in}
\alpha_{s}^{(0)}=\frac{6}{\beta}\,\frac{1}{4\pi}\ ,
\label{eq:boostedc} \eeq $\alpha_s^{(0)}$ is the bare strong
fine-structure constant and ${\cal P}$ is the average value of the
plaquette. We perform our simulation at $\beta=6.0$ for which
${\cal P}=0.59370$ and $\bar \alpha_{s}=0.13404$. This is the
value of the coupling which we use in our numerical estimates of
the renormalization constants. For example, at $\mu=2.0$~GeV we
obtain
 \bea \label{eq:zplusrin}
 {\cal Z}^{RI-MOM}_{+} = 0.600 \qquad
\textrm{and}\qquad
 \hat{\cal Z}^{RI-MOM} = \l ( \begin{array}{cc}
         0.631 &  -0.135 \\
          -0.025 & 0.302 \\
        \end{array}\r ) \ .
\eea
The numerical estimates of the renormalization constants which we
obtain using the boosted coupling in eq.~(\ref{eq:boostedc}) are
in reasonable agreement with those obtained non-perturbatively:
compare, for example, the results in eqs.~(\ref{eq:zplusrin})
with those in eqs.~(\ref{eq:zplusrinp}) and
(\ref{eq:z78rinp}) below. However, the one-loop corrections are so
large that we would not have trusted the values in
eqs.~(\ref{eq:zplusrin}) had we not also
evaluated the renormalization constants non-perturbatively (see
section \ref{subs:np}). On the other hand, consistency with the
perturbative calculation gives us confidence that the large
deviations from unity (or the unit matrix) found
non-perturbatively for the renormalization constants are genuine
and not simply a consequence of lattice systematic errors.
\subsection{Non-perturbative matching}
\label{subs:np}
The procedure necessary to compute the renormalization constants
of $\Delta I=3/2$ four-quark operators non-perturbatively using
the RI/MOM method has been presented in detail in
ref.~\cite{donini_4q_np_renorm} following the general philosophy
of ref.~\cite{rome_np_renorm}. Recently, an extensive analysis of
the renormalization constants of both bilinear and four-quark
operators for the non-perturbatively ${\cal O}(a)$-improved Wilson
action in the quenched approximation has been presented in
ref.~\cite{Becirevic:2004ny}. In that study, four different values
of the lattice coupling were considered: $\beta= 6.0$, $6.2$,
$6.4$ and $6.45$. For $\beta=6.0$, two lattice volumes were used:
$16^3\times 52$, with $500$ gluon configurations, and $24^3\times
64$, with $340$ configurations. Since the data for the large
volume and this work's data set 1 are exactly the same, here we
will briefly outline the renormalization procedure and refer the
reader to ref.~\cite{Becirevic:2004ny} for full details.

The non-perturbative determination of renormalization constants
with the RI/MOM method is based on the numerical evaluation, in
momentum space, of correlation functions of the relevant operators
between external quark and gluon states. For the operators ${\cal
O}_{4}$, ${\cal O}_{7}^{3/2}$ and ${\cal O}_{8}^{3/2}$, we compute
the Green functions $G_{i}(p)$ with $i=4,7,8$ (for clarity, we
suppress colour and spinor indices)
\begin{equation}\label{eq:unamp}
G_{i}(p)=\int\, \prod_{k=1}^{4}\, d^{4} x_{k}\;
e^{-i\, p\cdot (x_{1}\, -\, x_{2}\, +\, x_{3}\, -\, x_{4})}\,
\langle\,\psi_1(x_1)\,\bar\psi_2(x_2)\,\bar{\cal
O}_i(0)\, \psi_3(x_3)\,\bar\psi_4(x_4)\,\rangle\ ,
\end{equation}
with all four external legs at equal virtuality $p^{2}=\mu^{2}$.
The amputated Green functions, $\Lambda_i$, are defined by
\begin{equation}\label{eq:amp}
\Lambda_i(p)= S^{-1}(p)\, S^{-1}(p)\, G_i(p)\, S^{-1}(p)\,
S^{-1}(p)\,,
\end{equation}
where $S$ is the quark propagator
\begin{equation}\label{eq:quarkprop}
S(p)=\int\, d^{4} x\, e^{-i\, p\cdot x}\,
\langle\,\psi(x)\,\bar\psi(x)\,\rangle\,.
\end{equation}
The RI/MOM renormalization method consists in imposing that the
amputated Green functions, $\Lambda_i(p)$, computed in the chiral
limit, in a fixed gauge and at a given large scale
$p^{2}=\mu^{2}$, are equal to their tree-level values. In this
study we work in the Landau gauge. In practice, we use Dirac
projection operators, $P_{i}$, to implement the renormalization
conditions
\begin{equation}\label{eq:rencons}
\left. {\cal Z}_\psi^{-2}(\mu a)\, {\cal Z}_{ik}(\mu a)\,
\Tr\left[\Lambda_k(p)\, P_j\right]\right|_{p^2=\mu^2} =\delta_{ij}
\end{equation}
in the chiral limit. ${\cal Z}_\psi$ is the quark field
renormalization constant, which we calculate from the quark
propagator as in ref.~\cite{donini_4q_np_renorm}. The projection
operators satisfy the orthogonality relation
$\Tr[\Lambda_{i}^{(0)}\, P_{j}]\, =\, \delta_{ij}$ where
$\Lambda_{i}^{(0)}$ is the tree-level amputated Green function of
the operator ${\cal O}_{i}$. In this work, we use the projector
basis defined in ref.~\cite{donini_4q_np_renorm}.

For the RI/MOM method to be applicable, the renormalization scale
$\mu$  must be much larger than $\Lambda_{QCD}$, in order to be
able to use the Wilson coefficients computed in perturbation
theory, and much smaller than the inverse lattice spacing, to
avoid large discretization errors, so that formally \beq
\Lambda_{QCD} \ll \mu \ll a^{-1} \, .\eeq The precise range of
validity of the RI/MOM method for the operators studied in this
paper is investigated in the following subsection.

It is important to notice that the validity of the RI/MOM approach
relies on the fact that non-perturbative contributions to the
Green functions vanish asymptotically at large $p^{2}$. A possible
difficulty in the implementation of the above procedure comes from
the coupling of the operators to the Goldstone boson
\cite{Cudell:1998ic,Aoki:1999mr,Dawson:2000kh}. In
ref.~\cite{Becirevic:2004ny} it is shown that for parity violating
operators, such as the ones considered here, only a single pion
pole can be presented in the amputated projected  Green function,
while in the parity conserving case, the appearance of a double
pole is also possible. The Goldstone boson makes the chiral limit
of the amputated projected Green functions singular and hence the
extraction of the renormalization constants becomes difficult and
affected by large uncertainties. In order to subtract the
Goldstone pole before extrapolating to the chiral limit, we follow
the strategies described in detail in ref.~\cite{Becirevic:2004ny}
which are based in part in the method suggested in
ref.~\cite{Giusti:2000jr}. In ref.~\cite{Becirevic:2004ny} it is
demonstrated that the contribution of the Goldstone pole may
indeed affect the extrapolation of some renormalization constants
to the chiral limit but this problem can be overcome by
non-perturbatively subtracting the singular term. In this work, we
confirm the conclusions of ref.~\cite{Becirevic:2004ny} and in the
next subsection we present the numerical results for the
renormalization constants.

\subsection{Renormalization Group Behaviour of the Renormalization
Matrix}\label{subsec:rgbehaviour}
\begin{figure}
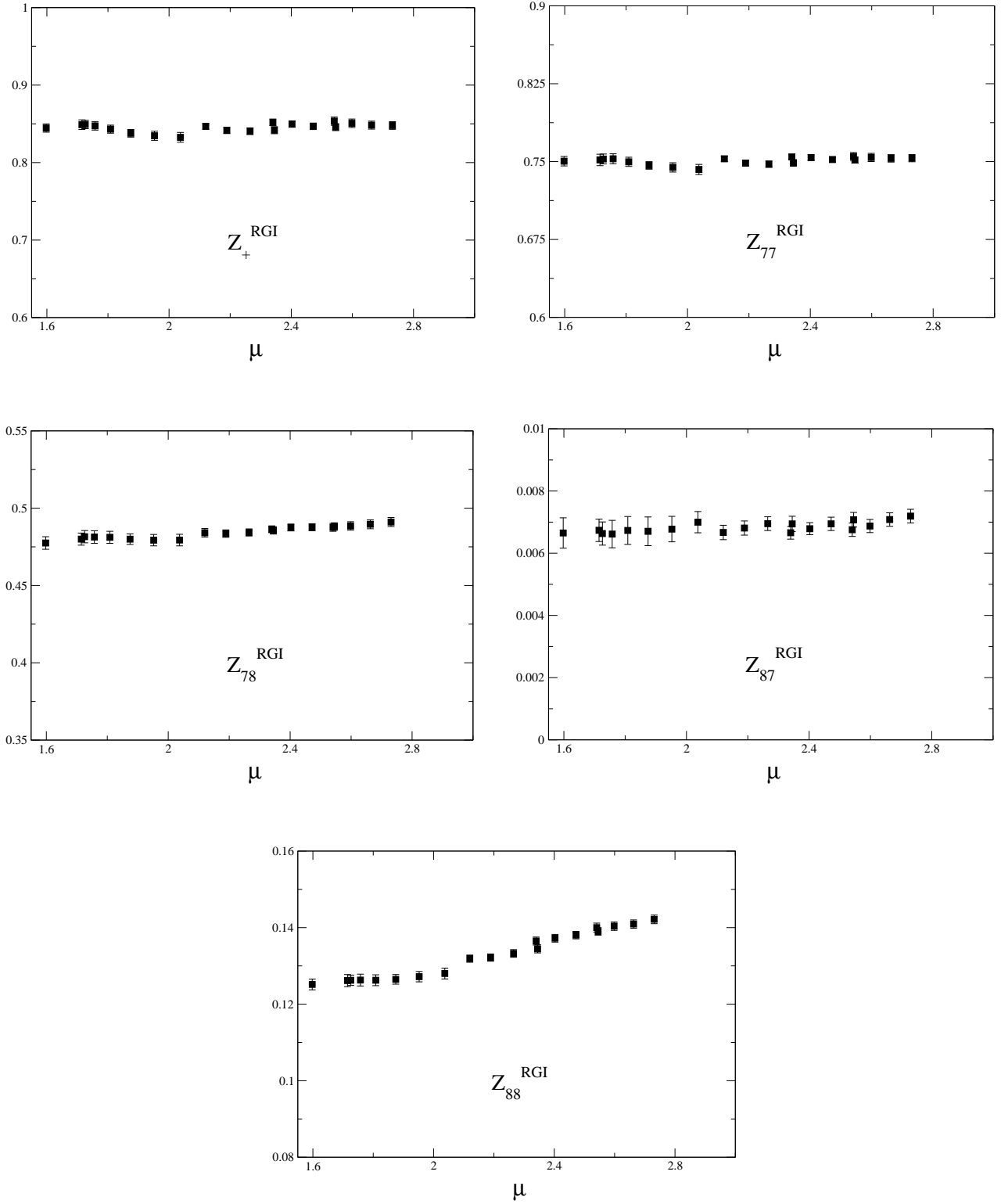

\begin{center}
\hspace*{-9cm}
\mbox{\epsfig{figure=figures/z+RGI.eps,angle=0,width=0.5\linewidth}\put(20,0)
{\epsfig{figure=figures/z77RGI.eps,angle=0,width=0.515\linewidth}}}
\\\vspace*{+1cm}\hspace*{-9cm}\mbox{\epsfig{figure=figures/z78RGI.eps,angle=0,width=0.505\linewidth}\put(20,0){\epsfig{figure=figures/z87RGI.eps,angle=0,
width=0.515\linewidth}}}\\
\vspace*{+1cm}\mbox{\epsfig{figure=figures/z88RGI.eps,angle=0,width=0.5\linewidth}}
\caption{\label{fig:RGI} $Z^{RGI}$ defined in
  eq.~(\ref{eq:appdRGI}), with $w_+$, $\hat w$ computed at NLO in
  perturbation theory and ${\cal Z}_+(\mu)$, $\hat{\cal Z}(\mu)$
  computed non-perturbatively in the RI-MOM scheme. The scale $\mu$ is
in $\gev$.}
\end{center}
\end{figure}

The renormalized operators, obtained either perturbatively or with
the non-perturbative method, should match the renormalization
dependence of the Wilson coefficients in order to obtain physical
amplitudes which are independent of $\mu$. Whereas this is
automatically enforced by a consistent use of the perturbative
renormalization constants at the next-to-leading order,  it is
useful to check whether the non-perturbative renormalization
constants also satisfy the expected renormalization group
behaviour. Following ref.~\cite{Becirevic:2004ny}, we thus define the renormalization group invariant
(RGI) renormalization constants by \beq \label{eq:appdRGI} {\cal
Z}_+^{RGI}(a) = w_+^{-1}(\mu){\cal Z}_+(\mu)\qquad\hat{\cal
Z}^{RGI}(a) = (\hat{w}^{-1})^T(\mu)\hat{\cal Z}(\mu)\;, \eeq where
$\hat w^{-1}(\mu)$ contains the scale dependence of the Wilson
coefficients. At NLO we write \beq\label{eq:wdefi}
(\hat{w}^{-1})^T(\mu)= \left[\alpha_s (\mu) \right]^{
      - \frac{\hat \gamma^{(0)}}{2\beta_{0}}}
      \left[1+{\alpha_s(\mu)\over 4\pi} \hat J^T\right]\; ,
\eeq with a similar expression for $w_+^{-1}(\mu)$ in terms of
$J_+$ and $\gamma^{(0)}_+$. The values of $J_+$ and $\hat{J}$ have
been computed at NLO in~\cite{Ciuchini:1998bw}. As mentioned
above, if we compute the quantities in eq.~(\ref{eq:appdRGI}) by
using the non-perturbatively determined renormalization constants
then the scale independence is not guaranteed. In
fig.~\ref{fig:RGI} we show the numerical results for the
quantities in eq.~(\ref{eq:appdRGI}) as functions of $\mu$.

Although the computed values of the $Z^{RGI}$ in
fig.~\ref{fig:RGI} are independent of $\mu$ to a reasonable
precision, we see that especially for ${\cal Z}_{88}$ the plateau
is worse than in the other cases. This could be due either to
lattice artefacts or to the possibility that higher order
contributions to the RG evolution may be significant. The detailed
study in ref.~\cite{Becirevic:2004ny}, using data from simulations
at four values of the lattice spacing, suggests that the lattice
artefacts are not large and interpret the deviation from scaling
as being due to N$^2$LO perturbative corrections. As emphasized in
ref.~\cite{Becirevic:2004ny} it would be desirable to control
better the scaling behaviour by evaluating the N$^2$LO anomalous
dimensions and/or performing a step-scaling analysis.

We extract ${\cal Z}_+^{RGI}$ and $\hat{\cal Z}^{RGI}$ by fitting
each of the plateaus in fig.~\ref{fig:RGI} to a constant in the
interval $2.1\gev\leq \mu\leq 2.4\gev$. We then use
eq.~(\ref{eq:appdRGI}) to obtain the value of the renormalization
constants at the required scale. A standard scale at which the
matrix elements are matched with the Wilson coefficients is
$\mu=2\gev$. At this value of $\mu$, the values of the
renormalization constants in the RI-MOM scheme are %
\bea \label{eq:zplusrinp}
 {\cal Z}^{RI-MOM}_{+} &=& 0.601(3)\left(^{+6}_{-9}\right) \\
 & & \nonumber \\
\label{eq:z78rinp}
 \hat{{\cal Z}}^{RI-MOM} &=& \l ( \begin{array}{cc}
\phantom{-}0.664(2)\left(^{+3}_{-8}\right) &-0.142(1)\left(^{+2}_{-2}\right) \\
-0.055(1)\left(^{+1}_{-2}\right) &\phantom{-}0.360(3)\left(^{+21}_{-21}\right) \\
\end{array}\r ) \,.
\eea The first error is statistical while the second one is the
systematic uncertainty estimated as explained in
sec.~\ref{sec:op_match_sys}. The conversion to the $\msbar$-NDR
and $\msbar$-HV renormalization schemes can be readily done by
using eqs.~(\ref{eq:matchRIMS1})-(\ref{eq:matchRIMS3}) (we take
$\alpha^{\msbar}_s(\mu=2 \gev)=0.1953$ obtained in the quenched
approximation at NLO with $\Lambda_{QCD}=0.250 \gev$). The
non-perturbatively obtained results in eqs.~(\ref{eq:zplusrinp})
and (\ref{eq:z78rinp}) are remarkably similar to the perturbative
results in eq.~(\ref{eq:zplusrin}) (the similarity is less
remarkable if mean field improvement is used to estimate the
higher order terms in perturbation theory). Moreover, we can
compare our non-perturbative results, computed with the data sets
1 and 2, with the ones obtained previously in
ref.~\cite{Becirevic:2004ny} at the same $\beta$ but using a
smaller volume and data set 1 only. In this study, a complete
basis of five parity odd $\Delta I=3/2$ operators was considered.
For our purposes we only need to consider three of these, namely
(in the notation of ref.~\cite{donini_4q_np_renorm}) ${\cal
Q}^+_1$, ${\cal Q}^+_2$, and ${\cal Q}^+_3$. In fact, with the
appropriate choice of flavour quantum numbers, there is the
following correspondence between the two notations: \bea & &{\cal
O}_4\equiv2{\cal Q}^+_1,\ \ {\cal O}^{3/2}_7\equiv2{\cal Q}^+_2,\
\
{\cal O}^{3/2}_8\equiv-4{\cal Q}^+_3\nonumber;\\
{\cal Z}_{+}\equiv{\cal Z}_{11},\ & &{\cal Z}_{77}\equiv{\cal
Z}_{22},\ \ {\cal Z}_{78}\equiv -\frac{1}{2}\, {\cal Z}_{23},\ \
{\cal Z}_{88}\equiv{\cal Z}_{33},\ \ {\cal Z}_{87}\equiv -2\,
{\cal Z}_{32}, \eea and the renormalization structure is therefore
the one indicated in eqs.~(\ref{eq:o4ren}) and (\ref{eq:o78ren}).
The values obtained in ref.~\cite{Becirevic:2004ny} were the
following, \beq\label{eq:zold}
\begin{array}{rccc}
{\cal Z}^{RI-MOM}_{+} &=& 0.608(4)(14) & 0.604(3)(5)\\
\hat{{\cal Z}}^{RI-MOM} &=& \l ( \begin{array}{cc}
\phantom{-}0.673(3)(10) &-0.139(1)(5) \\
-0.056(2)(2) &\phantom{-}0.392(3)(28) \\
\end{array}\r )&
\l ( \begin{array}{cc}
\phantom{-}0.666(3)(3) &-0.140(1)(2) \\
-0.056(2)(2) &\phantom{-}0.375(3)(14) \\
\end{array}\r )\\
\end{array}
\eeq where the first corresponds to the small volume $16^{3}\times
52$ and the second to data set 1. As can be seen by comparing
eqs.(\ref{eq:zplusrinp}) and (\ref{eq:z78rinp}) with
eq.(\ref{eq:zold}), the agreement is excellent, suggesting that
finite-volume effects are very small.

\section{The Analysis}
\label{sec:analysis}
In this section we discuss the analysis of our results, starting
with the extraction of the matrix elements. We then discuss the
determination of the shift of the energy of the two-pions as a
result of finite-volume effects and finally we explain our
procedure for the chiral extrapolation.

\subsection{Extraction of the Matrix Elements}
\label{sec:extrme}
We now explain the procedure we use to determine the matrix
elements in the SPQR kinematics \cite{Boucaud:2001mg, Lin:2002nq},
where one of the final-state pions has a non-zero spatial
momentum, and the other two mesons are at rest. In order to remove
the $I=1$ component in the two-pion state, we take the symmetric
combination of the two states
$|\pi^{+}({\vec{p}}^{\,})\pi^{0}(\vec{0})\ra$ and
$|\pi^{+}(\vec{0})\pi^{0}(\vec{p}^{\,})\ra$, where $|\vec{p\,}|=0$
or $2\pi/L$, i.e. we evaluate the following average:
\beq \label{eq:spqr_amp}
\ampl^{(i)}_{\mathrm{SPQR}}(M_{\pi},E_{\pi},M_{K}) = \frac{1}{2}
\left ( \la\pi^{+}(\vec{p}^{\,})
\pi^{0}(\vec{0})|\op_{i}|K^{+}(\vec{0})\ra
  + \la\pi^{+}(\vec{0})\pi^{0}(\vec{p}^{\,})|\op_{i}|K^{+}(\vec{0})\ra\right )
  ,
\eeq where $E_\pi$ is the energy of the corresponding pion and
$\op_i$ are the operators defined in
eqs.~(\ref{eq:o4_def})-(\ref{eq:o8_32_def}).

\begin{center}
\begin{figure}
\begin{center}
\hspace*{-8.7cm}
\mbox{\epsfig{figure=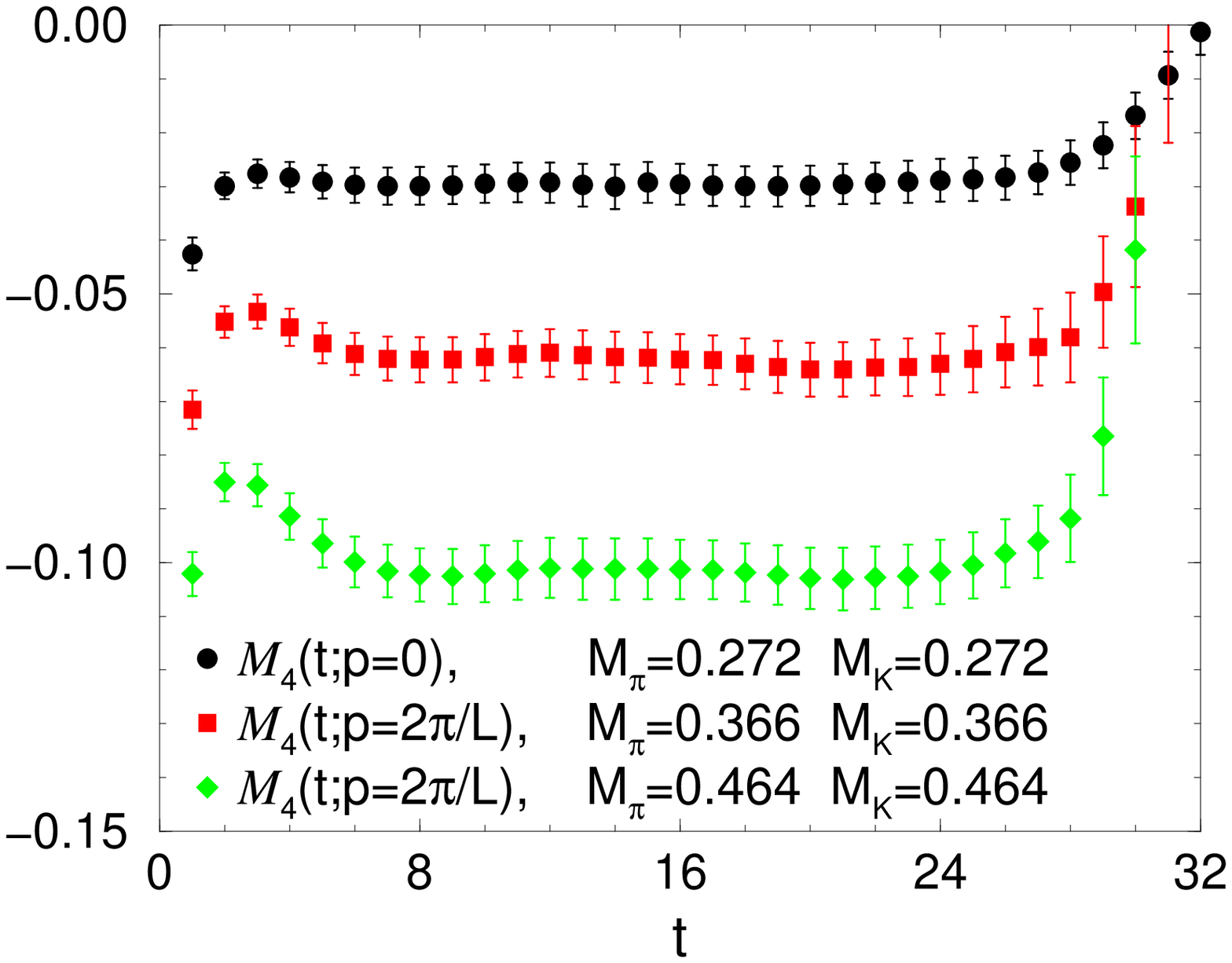,angle=0,width=0.525\linewidth}\put(15,3)
{\epsfig{figure=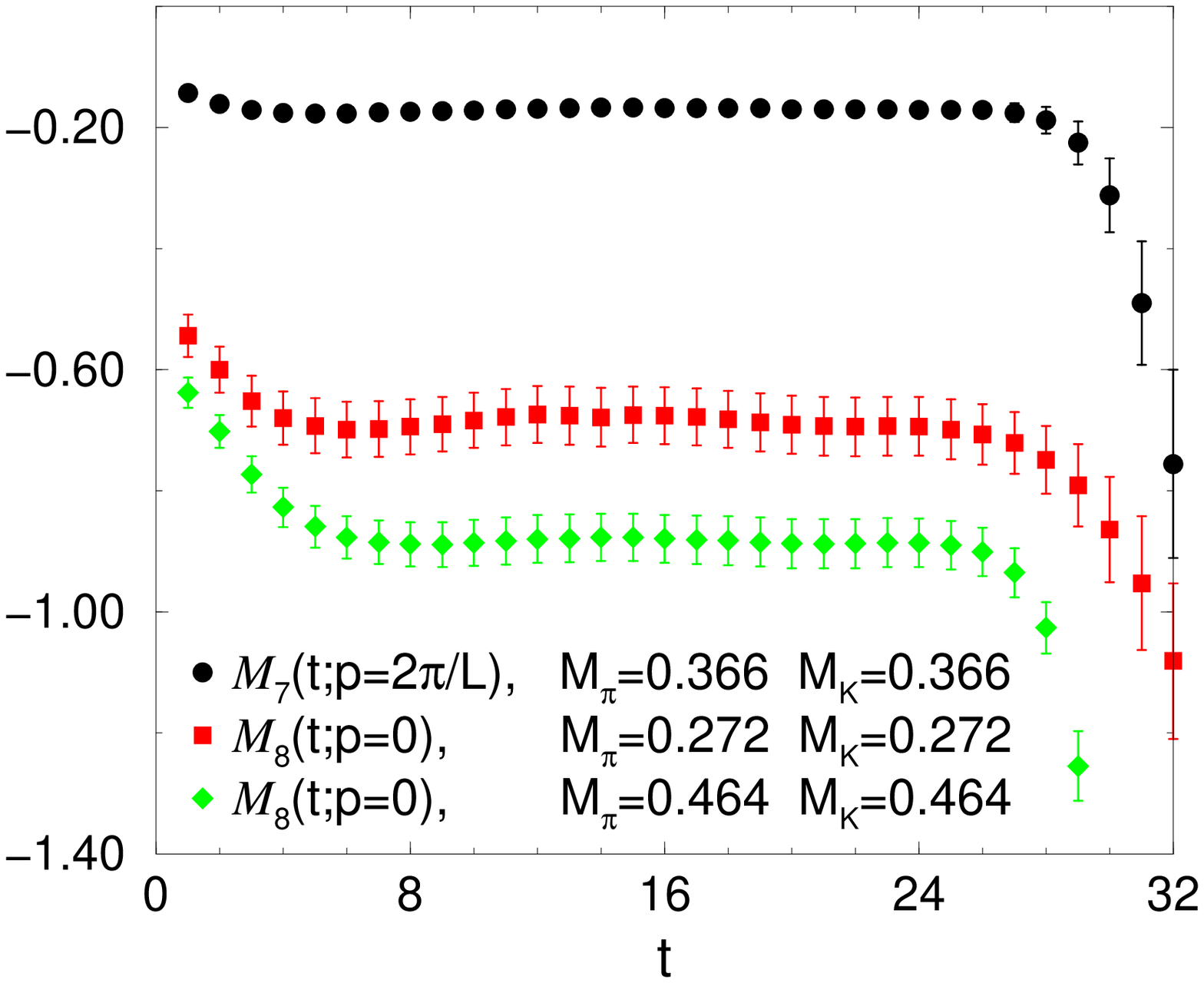,angle=0,width=0.49\linewidth}}}
\caption{\label{fig:Oplateaux} Plateaus for the extraction of the
form factors $a^{3}\bar{\ampl}^{(i)}_{\mathrm{SPQR}}$.}
\end{center}
\end{figure}
\end{center}

In Euclidean space, up to NLO precision in the chiral expansion
and neglecting finite-volume effects, the magnitude of
$\ampl^{(i)}_{\mathrm{SPQR}}$ can be extracted from the ratios of
correlation functions
\beq
\label{eq:ratio}
  \frac{1}{2} \l ( \frac{C^{(i)}_{+}(t,\vec{p}^{\,}) +
      C^{(i)}_{0}(t,\vec{p}^{\,})}{C_{K}(t_{\textrm{\scriptsize{fix}}},\vec{0})
       C_{\pi}(t_{\textrm{\scriptsize{fix}}},\vec{0})C_{\pi}(t,\vec{p}^{\,})} \r )
   (a^{6}Z_{\pi}^{2}Z_{K})
      \stackrel{\mathrm{large}\mbox{ }t}{\longrightarrow}
 a^{3}\bar{\ampl}^{(i)}_{\mathrm{SPQR}} ,
\eeq
where $C^{(i)}_{+}$ and $C^{(i)}_{0}$ are defined in eq.(\ref{eq:4q_correlator})
and $\bar{\ampl}^{(i)}_{\mathrm{SPQR}}$ represents the magnitude
of the matrix element of the bare lattice operator $\bar{\op}_{i}$
on the right-hand side of eq.~(\ref{eq:spqr_amp}). $Z_{P}=\la
0|\phi_P(0)|P\ra$ is the wave-function renormalization of the
pseudoscalar meson state $P$, and can be obtained from the fit to
\beq \label{eq:ps_fit} C_{P}(t,\vec{0}) =
\frac{(a^2Z_{P})^{2}}{aM_{P}}\
            {\mathrm{e}}^{-aM_{P}\frac{T}{2}}\ {\mathrm{cosh}}\l [
            aM_{P}\l ( \frac{T}{2} - t\r )\r ],
\eeq
at large $t$, where $T$ is the number of time slices (T=64 in this
study). Fig.~\ref{fig:Oplateaux} shows some of the plateaus for
$a^{3}\bar{\ampl}^{(i)}_{\mathrm{SPQR}}$.
The value of the pseudoscalar decay constant $f_{P}$ corresponding
to the (non-degenerate) mass $M_P$ is obtained via the
one-parameter fit (using $Z_{P}$ and $M_{P}$ obtained from eq.
(\ref{eq:ps_fit}))
\beq
 \left ( \frac{C_{PA}(t,\vec{0})}{C_{PP}(t,\vec{0})} \right )
 \times
 \left ( \frac{Z_{P}}{M_{P}}\right )=
 \left ( \frac{f_{P}a}{Z_{A}}
 \right )\times {\mathrm{tanh}}\left [ a M_{P}
 \left ( \frac{T}{2}-t\right )\right ] ,
\eeq
at large $t$. In order to obtain $f$ we neglect the effects of
$SU(3)$ breaking and we extrapolate quadratically on the smallest
5 masses. We obtain $f=0.139(5)\gev$ from data set 1 and
$f=0.135(4)\gev$ from set 2. The renormalization constant $Z_{A}$
is determined non-perturbatively, as discussed in
sec.~\ref{sec:op_match}.

Eq.~(\ref{eq:ratio}) corresponds to the case in which NLO $\chi$PT
is applicable and in which finite-volume effects are neglected.
More generally we use the finite-volume formalism of
refs.~\cite{Lellouch:2000pv,Lin:2001ek} to write:
\begin{eqnarray}
  \frac{1}{2} \l ( \frac{C^{(i)}_{+}(t,\vec{p}) +
      C^{(i)}_{0}(t,\vec{p})}{C_{K}(t_{\textrm{\scriptsize{fix}}},\vec{0})
       C_{\pi}(t_{\textrm{\scriptsize{fix}}},\vec{0})C_{\pi}(t,\vec{p})} \r )
   (a^{3}Z_{\pi}^{2}Z_{K})&
      \stackrel{\mathrm{large}\mbox{
      }t}{\longrightarrow}&\nonumber\\ &&\hspace{-2in}
 \left\{a^{3}|\bar{\ampl}^{(i)}_{\mathrm{SPQR}}|\cos \delta^{I=2}(W) +
      O\left(\frac{1}{L}\right)\right\}e^{-\Delta W t_{\textrm{\scriptsize{fix}}}} ,
\label{eq:ratio1}\end{eqnarray} where $W$ is the two pion energy
in a finite volume so that $\Delta W\equiv W-E_\pi-M_\pi$ and
$\delta^{I=2}(W)$ is the $I=2$ two-pion phase shift. We now
discuss the finite-volume effects in more detail.

\subsection{Two Pion Energy Shift in a Finite Volume}
\label{sec:deltaW}

In this section we discuss the extraction of the energy shift of
the two-pion state with the SPQR kinematics, $\Delta W\equiv
W-E_\pi-M_\pi$, using a data set of 750 configurations (of which
set 1 is the subset on which the four-point functions were also
computed). We consider pions of three masses, each composed of a
degenerate quark-antiquark pair. The masses of the pions in
lattice units are $M_\pi=$ 0.4438(7), 0.3590(8) and 0.2557(13)
corresponding to hopping parameters $\kappa=0.13300,$ 0.13376 and
0.13449 respectively.
\begin{figure}[!b]
\epsfig{figure=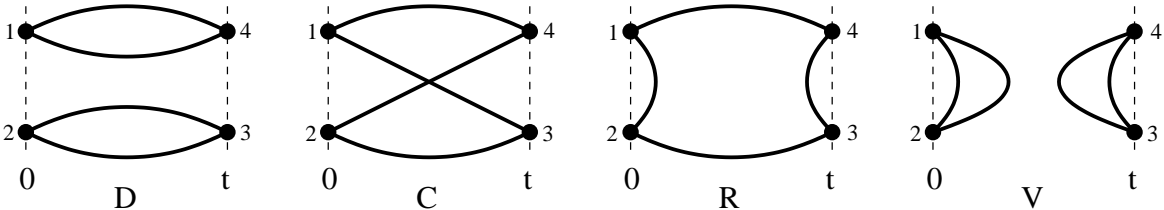,angle=0,width=1.0\linewidth}
\caption{\sl Feynman diagrams which appear in the computation of
the most generic four pion correlation function as
$G_{4\pi}(t,\vettp)$ in the $SU(2)$ symmetric
limit.\label{fig:diagI2}}
\end{figure}

Consider the correlation function
\begin{eqnarray}
G_{4\pi}(t,\vettp)&=&\frac{1}{2}\Big(\left\<P_{\pi^{+}}(t,\vettp)
P_{\pi^{0}}(t,\vetto)\phi_{\pi^{+}}^\dag(\vetto,0)\phi_{\pi^{0}}^\dag(\vetto,0)\right\>
\nonumber\\
&&\hspace{0.3in}+\left\<P_{\pi^{+}}(t,\vetto)
P_{\pi^{0}}(t,\vettp)\phi_{\pi^{+}}^\dag(\vetto,0)\phi_{\pi^{0}}^\dag(\vetto,0)\right\>
\Big) \label{eq:4pi} \end{eqnarray}where the interpolating
operators in momentum and position space, $P_\pi$ and $\phi_\pi$
respectively, are defined in eq.~(\ref{eq:interpol_op}). We have
symmetrized over the two interpolating operators at fixed momentum
$P_{\pi^{+}}$ and $P_{\pi^{0}}$ in order to obtain a pure $I=2$
state. The most general four-pion correlation function may be
represented in terms of the four diagrams in
fig.~\ref{fig:diagI2}. Moreover, in the $SU(2)$ symmetric limit,
since the two interpolating operators at $t=0$ are at the same
spatial point, the symmetrization performed in eq.~(\ref{eq:4pi})
is automatic. The $I=2$ correlation function in eq.~(\ref{eq:4pi})
is given by the combination $D-C$ of diagrams in
fig.~\ref{fig:diagI2}, while the corresponding $I=0$ correlator is
given by the combination $D+C/2-3R+3/2V$. Here we only consider
the $I=2$ channel, which is the relevant one for \ditre
transitions.

To determine the two-pion energy shift $\Delta W$, we compute the
ratio \beq R_{4\pi}(t;\vettp)\equiv \frac{G_{4\pi}(t;\vettp)
}{C_{\pi}(t,\vettp)C_{\pi}(t,\vetto)}\,,\ \label{eq:ratio4pidef}
\eeq where $C_{\pi}$ is defined in eq.~(\ref{eq:2pts_pp}), and, in
order to isolate the lightest states, we consider the behaviour of
the correlation functions in the region $T\gg t\gg0$.

On a lattice with periodic boundary condition in the temporal
direction $G_{4\pi}$ has two types of contribution; either a
two-pion state propagates between 0 and $t$ (or between $t$ and
$T$) or there is a single pion propagating between 0 and $t$ and
another between $t$ and $T$. In the region $T\gg t\gg0$ the ratio
is therefore given by:\bea
R_{4\pi}(t,\vettp)&\longrightarrow&\frac{X_1\big(e^{-W
t}+e^{-W(T-t)}\big) + X_2\big( e^{-E_\pi t} e^{-M_\pi(T-t)} +
e^{-M_\pi t} e^{-E_\pi(T-t)}\big)}{X_3 \big(e^{-M_\pi
t}+e^{-M_\pi(T-t)}\big)\big(e^{-E_\pi
t}+e^{-E_\pi(T-t)}\big)}\nn\\
&&\hspace{-0.7in}=\frac{X_1
e^{-W\frac{T}{2}}\cosh\big(W(t-\frac{T}{2}) \big) +
X_2e^{-\frac{(M_\pi+E_\pi)}{2}T}\cosh
\big((E_\pi-M_\pi)(t-\frac{T}{2}))}{2\;X_3e^{-\frac{(M_\pi+E_\pi)}{2}
T}\cosh\big(M_\pi (t-\frac{T}{2})\big)\cosh\big(E_\pi
(t-\frac{T}{2})\big)} \label{eq:ratio4pi} \eea where the constants
$X_{1,2,3}$ are independent of time but do depend on the finite
volume energies of the $|\pi\>$ and $|\pi\pi\>$ states. $E_\pi$ is
the energy of a pion with momentum $\vettp$. We have not displayed
the contribution of the excited states.
\begin{figure}[!b]
\mbox{\hspace*{-0.3cm}
\epsfig{figure=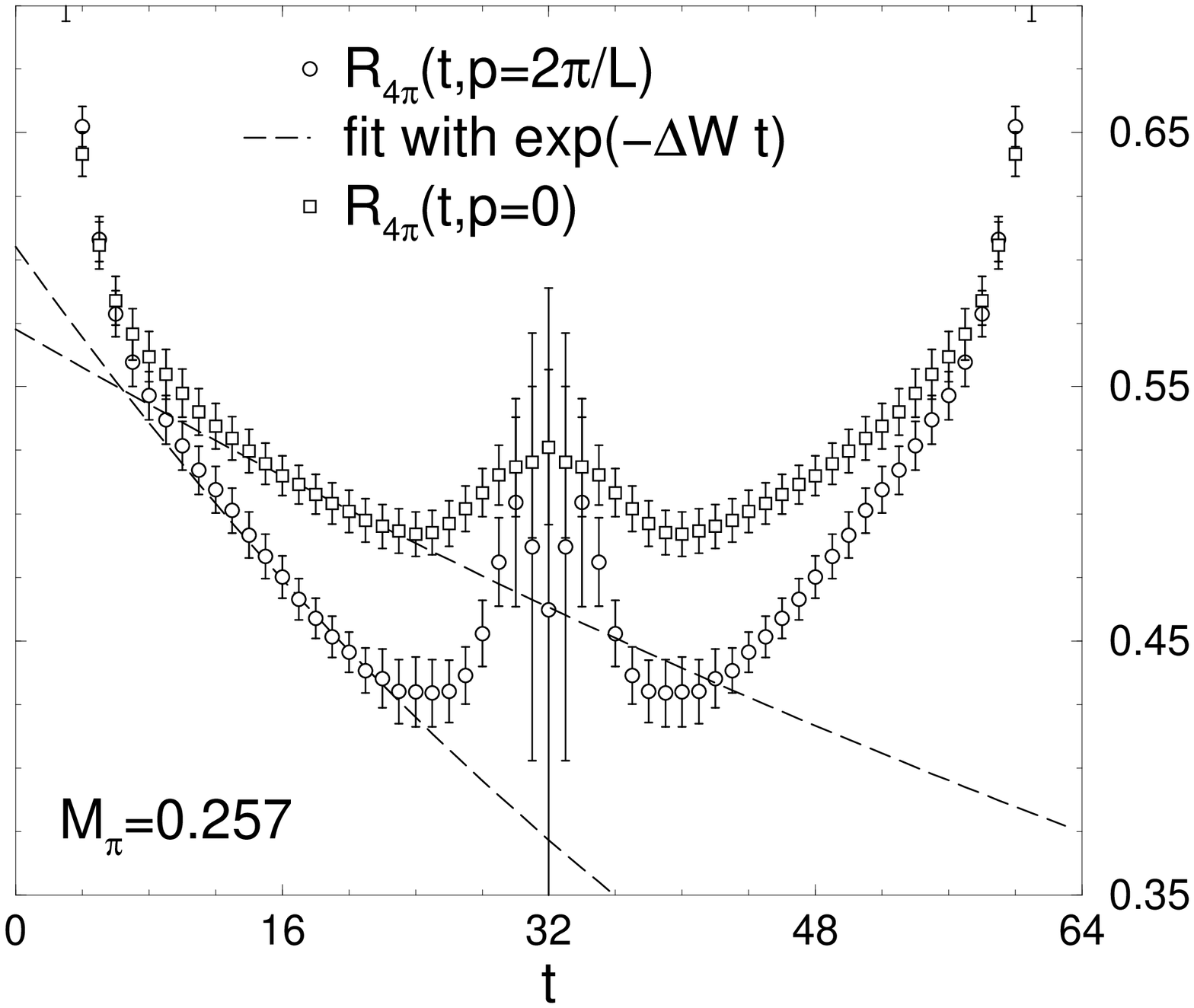,angle=-90,width=0.50\linewidth}
\put(8,-3){\epsfig{figure=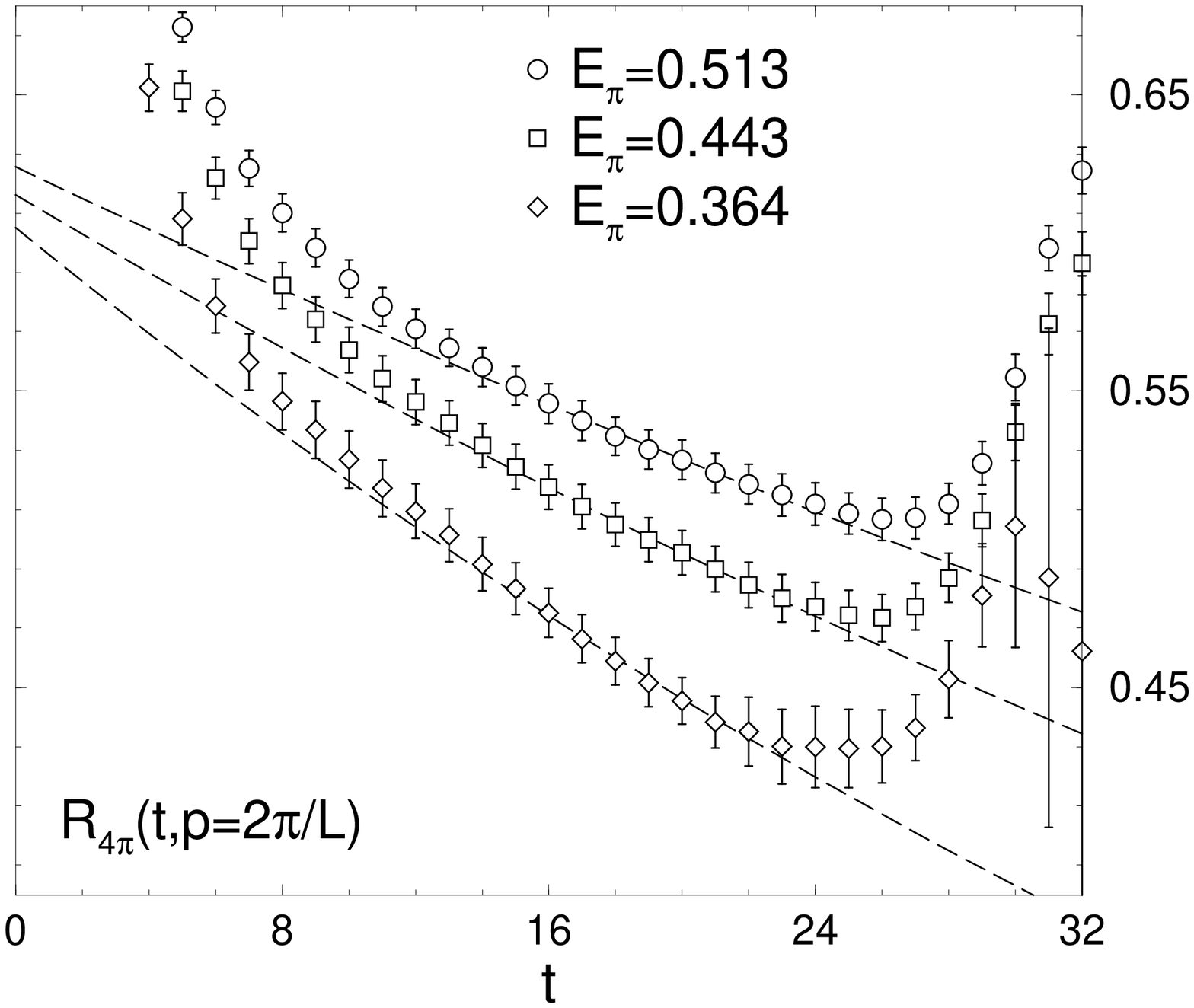,angle=-90,width=0.50\linewidth}}}
\vspace*{-0.4cm} \caption{\sl (a) $R_{4\pi}(t,\vettp)$ for
$aM_\pi=0.257$ and for both momenta, on the whole lattice. (b)
$R_{4\pi}(t,\vettp_{\textrm{\scriptsize{min}}})$ for the three
masses of the pion.\label{fig:ratioW1}}
\end{figure}

In fig.~\ref{fig:ratioW1}\,(a) we plot $R_{4\pi}(t,\vetto)$ and
$R_{4\pi}(t,\vettp_{\textrm{\scriptsize{min}}})$ as a function of
$t$ for the lightest quark mass (corresponding to
$\kappa=0.13449$). $|\vettp_{\textrm{\scriptsize{min}}}|=2\pi/L$
and we average over the six equivalent directions. In
fig.~\ref{fig:ratioW1}\,(b) we plot
$R_{4\pi}(t,\vettp_{\textrm{\scriptsize{min}}})$ for all three
pion masses. In order to determine $\Delta W$ we have to fit the
$R_{4\pi}(t,\vettp)$'s in a range of $t$ such that the first term
in the numerator of eq.~(\ref{eq:ratio4pi}) dominates over the
second. This condition is in addition to the usual requirement
that the time intervals are sufficiently large for the
contributions from excited states to be suppressed. In practice we
extract the energy shifts by fitting $R_{4\pi}(t,\vettp)$ to $A
\exp(-\Delta W t)$ in the range $t\in[14-15,\,22-23]$
(symmetrizing the contributions from the correlation functions
over times in the forward and backward halves of the lattice),
with the exact choice of interval depending on the mass and
momentum. In order to estimate the systematic error, we vary the
limits of the time interval used for the fit and look at the
spread of the values thus obtained. The results are reported in
tab.~\ref{tab:deltae} (preliminary results were presented in
ref.\,\cite{Boucaud:2001tx}).

\begin{table}[!t]
\centering
\begin{tabular}{cccc}
\hline\hline
$M_\pi$ & 0.4438(7) & 0.3590(8) & 0.2557(13)\\
\hline
$\Delta W_{\vettp=0}$& 0.0054(6)(7) & 0.0062(7)(7) & 0.0066(9)(7)\\
$\Delta W_{\vettp=\frac{2\pi}{L}}$& 0.0086(8)(10) & 0.0109(11)(15) &
0.0152(27)(15)\\
$\Delta W_L$& 0.0064(3) & 0.0072(3) & 0.0083(5)\\
$\Delta W_{BG}$& 0.0052(2) & 0.0060(2) & 0.0071(3)\\
\hline\hline
\end{tabular}
\caption{\sl Numerical results ($\Delta W_{\vettp=0}$ and $\Delta
W_{\vettp=\frac{2\pi}{L}}$) and theoretical predictions ($\Delta
W_L$ and $\Delta W_{BG}$ for $\vettp=0$, defined in eqs.~(\ref{eq:lushift2}) and
(\ref{eq:berngolt2})). For the numerical results, which are given
in lattice units, we quote statistical and systematic
uncertainties in that order.\label{tab:deltae}}
\end{table}

In tab.~\ref{tab:deltae} we also present theoretical predictions
for the energy shift obtained using the L\"uscher formula and
tree-level $\chi$PT in full QCD ($\Delta
W_L$)~\cite{ml1,ml2,ml3,ml4} and by using one-loop quenched
$\chi$PT~($\Delta W_{BG}$)~\cite{Bernard:1996ez}. In both cases
the predictions are for $\vettp=0$. In spite of reservations that
our pion masses may be too large for $\chi$PT to be applicable,
the agreement of our lattice results with the theoretical
predictions from one-loop quenched $\chi$PT is very good and
reassuring. On the other hand, we observe discrepancies between
our results and the predictions of Luscher's formula (which
however is obtained in full QCD). We now outline the determination
of $\Delta W_L$ and $\Delta W_{BG}$.

The L\"uscher formula for the energy shift gives \bea
\label{eq:lushift2} && \Delta W_L = W-2M_\pi = -\frac{4\pi
a_0}{M_\pi L^3}\left\{1+c_1
\frac{a_0}{L}+c_2\frac{a_0^2}{L^2}\right\}+O(L^{-6})\,,\\&&\nn \\
&&\textrm{where}\qquad c_1=-2.837297,\qquad c_2=6.375183 \eea and
$a_0$ is the scattering length in the $s$-wave, $I=2$ channel. At
tree level in full $\chi$PT \beq a_0 = -\frac{M_\pi}{8\pi
f_\pi^2}\;. \label{eq:scatleng2} \eeq  We use the values of
$M_\pi$ obtained in our simulation and the corresponding values of
$f_\pi(M_\pi)$.

$\Delta W_{BG}$ is obtained by using one-loop quenched ChPT: \beq
\Delta W_{BG} = \Delta W^{\textrm{\scriptsize{tree}}}+\Delta
W^{\textrm{\scriptsize{1-loop}}}= \frac{1}{2 f_\pi^2 L^3} +
M_\pi\left( B_2(M_\pi L) \delta^2 + A_2(M_\pi L) \delta \epsilon +
O\Big(\frac{\epsilon^2}{(M_\pi L)^3}\Big)\right)
\label{eq:berngolt2} \eeq where the first term corresponds to the
leading term in the L\"uscher formula after substitution of the
expression for the scattering length in eq.~(\ref{eq:scatleng2}),
while the second term comes from one-loop diagrams. In the region
of large $M_\pi L$, $A_2(M_\pi L)$ and $B_2(M_\pi L)$  can be
approximated by \bea A_2(M_\pi L)\simeq \frac{1}{\pi}
\left(\frac{2\pi}{M_\pi L}\right)^3 \qquad\textrm{and}\qquad
B_2(M_\pi L)\simeq -\frac{1}{6 \pi} \left(\frac{2\pi}{M_\pi
L}\right)^3\;. \eea Finally \beq \delta\equiv
\frac{m_0^2/3}{8\pi^2 f_\pi^2}\qquad\textrm{and}\qquad
\epsilon\equiv \frac{M_\pi^2}{16\pi^2 f_\pi^2}\,,
\label{eq:deltaepsilon} \eeq where $m_0$ is the $\eta'$ mass.
Taking the full QCD estimate $m_0^2/3=(500\mev)^2$ would give
$\delta\sim 0.18$. We vary $\delta$ between $0.08$ and $0.18$
(which is approximately the range of values typically obtained in
quenched simulations). In this interval, however, the one-loop
contribution is small for the parameters of our simulation, and we
observe a very mild dependence of $\Delta W_{BG}$ on $\delta$
which is included in the quoted uncertainty. In
tab.~\ref{tab:deltae} we present the results obtained for
$\delta=0.10$.

In our data set 2 for $K\to\pi\pi$ decays we have not evaluated
the correlation function $G_{4\pi}$ and therefore cannot determine
$\Delta W$ at the corresponding pion masses directly. However, as
shown in tab.~\ref{tab:ps_mass} the pion masses in this data set
are very close to those in set 1 and we therefore obtain the
energy shifts at these masses by interpolating or extrapolating
the results obtained in set 1. We find that for the pion masses
$aM_\pi=$ 0.4642, 0.3663 and 0.2716, the corresponding $a\Delta
W_{\vettp=0}=$ 0.0052, 0.0061, 0.0065 and $a\Delta
W_{\vettp=\frac{2\pi}{L}}=$ 0.0080, 0.0107, 0.0146.

Finally we use the extracted values of $\Delta W_{\vettp}$ for
SPQR kinematics to correct the ratio on the left-hand side of
eq.~(\ref{eq:ratio1}) for the finite-volume effects $e^{-\Delta W
t_{\textrm{\scriptsize{fix}}}}$. Results for the matrix elements
of bare operators obtained in this way with $\cos(\delta)=1$ are
reported in tab.~\ref{tab:amplitudes}. They still contain
finite-volume effects, and in order to remove these one would need
to know the Lellouch-L\"uscher factors for SPQR kinematics. These
are currently unknown and we estimate the corresponding errors in
sec.\,\ref{sec:phase_shift_error}.

\begin{table}
\begin{center}
\begin{tabular}{c|cc|cc|cc}
\hline \hline $\#$&$\bar{\mathcal{M}}^{(4)}_{\rm SPQR}(\vetto)$&
$\bar{\mathcal{M}}^{(4)}_{\rm
SPQR}(\vettp_{\textrm{\scriptsize{min}}})$&
$\bar{\mathcal{M}}^{(7)}_{\rm SPQR}(\vetto)$&
$\bar{\mathcal{M}}^{(7)}_{\rm SPQR}(\vettp_{\textrm{\scriptsize{min}}})$
&$\bar{\mathcal{M}}^{(8)}_{\rm SPQR}(\vetto)$&
$\bar{\mathcal{M}}^{(8)}_{\rm SPQR}(\vettp_{\textrm{\scriptsize{min}}})$\\
\hline
\hline
\multicolumn{7}{c}{Set 1 (340 configurations)}\\
\hline
\hline
1a&0.0236(40)&0.0294(50)&0.199(20)&0.159(17)&0.695(71)&0.613(64)\\
2a&0.0298(42)&0.0384(68)&0.199(21)&0.157(15)&0.704(71)&0.620(65)\\
3a&0.0354(48)&0.0454(77)&0.202(22)&0.157(15)&0.725(72)&0.631(63)\\
\hline
1b&0.0521(55)&0.0570(66)&0.194(16)&0.167(14)&0.747(60)&0.685(56)\\
2b&0.0604(61)&0.0657(70)&0.195(15)&0.167(13)&0.767(59)&0.698(56)\\
3b&0.0712(73)&0.0769(82)&0.191(15)&0.162(13)&0.777(59)&0.702(56)\\
\hline
1c&0.0846(79)&0.0864(84)&0.191(14)&0.167(13)&0.815(61)&0.754(57)\\
2c&0.0952(87)&0.0972(92)&0.191(14)&0.167(12)&0.834(60)&0.771(58)\\
3c&0.1020(94)&0.1044(99)&0.187(14)&0.164(12)&0.834(60)&0.770(59)\\
\hline
\hline
\multicolumn{7}{c}{Set 2 (480 configurations)}\\
\hline
\hline
1a&0.0297(38)&0.0398(57)&0.187(14)&0.163(11)&0.687(47)&0.647(40)\\
2a&0.0320(37)&0.0428(57)&0.188(12)&0.166(11)&0.701(45)&0.661(39)\\
3a&0.0347(37)&0.0457(56)&0.190(12)&0.167(10)&0.713(44)&0.669(37)\\
4a&0.0374(39)&0.0487(56)&0.192(11)&0.170(10)&0.723(43)&0.679(36)\\
5a&0.0397(40)&0.0513(56)&0.193(11)&0.171(10)&0.729(42)&0.685(35)\\
6a&0.0416(42)&0.0538(57)&0.193(11)&0.172(10)&0.732(42)&0.691(35)\\
7a&0.0449(44)&0.0576(59)&0.194(11)&0.174(10)&0.740(42)&0.697(35)\\
\hline
1b&0.0583(43)&0.0636(50)&0.190(12)&0.170(10)&0.770(41)&0.718(35)\\
2b&0.0557(43)&0.0610(51)&0.190(11)&0.167(10)&0.761(42)&0.703(34)\\
3b&0.0615(44)&0.0671(50)&0.190(11)&0.171(10)&0.777(40)&0.730(35)\\
4b&0.0654(46)&0.0713(51)&0.192(10)&0.173(10)&0.789(39)&0.746(34)\\
5b&0.0681(47)&0.0743(51)&0.192(10)&0.174(10)&0.795(39)&0.751(34)\\
6b&0.0707(47)&0.0771(52)&0.192(10)&0.174(10)&0.800(38)&0.756(35)\\
\hline
1c&0.1012(54)&0.1025(57)&0.195(11)&0.176(10)&0.884(40)&0.833(37)\\
2c&0.0944(53)&0.0949(57)&0.194(12)&0.174(10)&0.868(42)&0.807(37)\\
3c&0.0973(53)&0.0982(57)&0.194(11)&0.175(10)&0.874(41)&0.819(37)\\
4c&0.1059(55)&0.1077(58)&0.195(10)&0.178(10)&0.896(39)&0.847(37)\\
5c&0.1088(55)&0.1107(57)&0.196(10)&0.179(10)&0.902(38)&0.855(37)\\
\hline
\hline
\end{tabular}
\end{center}
\caption{\label{tab:amplitudes}Numerical values of the bare
amplitudes $\bar{\ampl}^{(4,7,8)}_{\rm
SPQR}(\vettp)\equiv\bar{\ampl}^{(4,7,8)}_{\rm
SPQR}(M_{\pi},E_{\pi}(\vettp),M_{K})$ as obtained from the ratio
on the l.h.s. of eq.\,(\ref{eq:ratio1}) multiplied by the FV
correction $e^{\Delta W t_{\textrm{\scriptsize{fix}}}}$. $M_\pi$
assumes the values 1a, 1b, 1c and $M_K$ respectively
$\{1a,2a,\ldots\}$, $\{1b,2b,\ldots\}$, $\{1c,2c,\ldots\}$.}
\end{table}

\subsection{Chiral Extrapolations and the Numerical Results}
\label{sec:ana_details}

We start by fitting our lattice data to the formulae of the NLO
chiral expansion, including the chiral logarithms obtained in ref.
\cite{Lin:2002nq}, determine the low energy constants and use
these to extrapolate to the physical point. When applicable, we
use the numerical values for the couplings $\alpha = 0.1$ and
$m_{0}=0.5\gev$ in the terms
$\alpha(\partial_\mu\Phi_0)(\partial^\mu\Phi_0)-m^2_0\Phi^2_0$ of
the quenched chiral Lagrangian, where $\Phi_0$ is the
super-$\eta^\prime$ field. The results are not very sensitive to
the variation of $\alpha$ and $m_{0}$ within a reasonable range.
The fits are poor, which may not be surprising since we are trying
to use the chiral expansion in a range of masses and energies
which are rather large ($[0.52,1.5]\gev$). To improve the
situation we have also performed fits in restricted intervals,
introducing an upper cut-off for the masses and energies (we vary
this cut-off down to a value of 0.8\,GeV). Since the validity of
quenched $\chi$PT (q$\chi$PT) may be questioned even in the
restricted interval we have also performed fits using only the
polynomial terms of q$\chi$PT (i.e. the NLO expressions from
q$\chi$PT but without the logarithms) which have the same
functional form as those in $\chi$PT of full QCD.
\begin{figure}
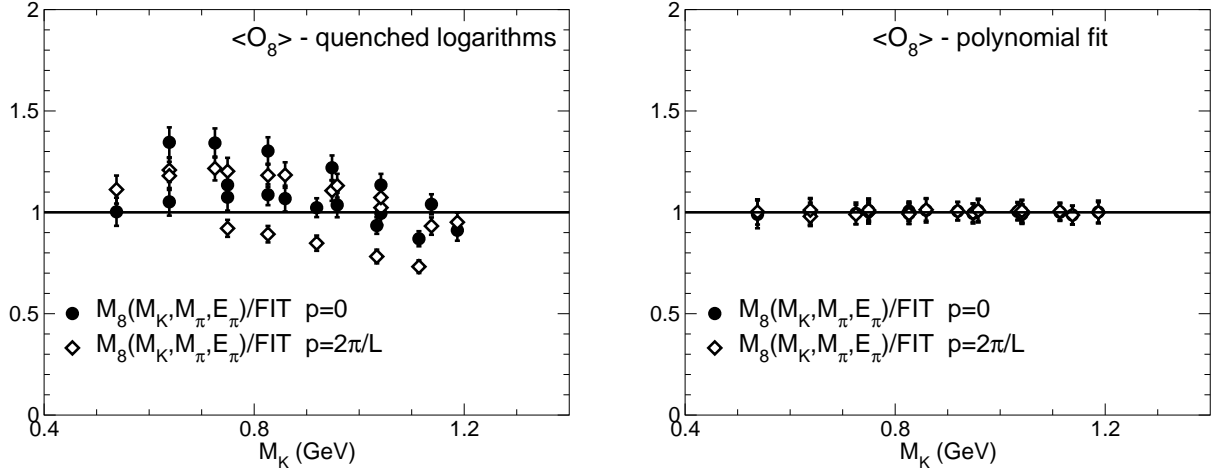

\begin{center}
\hspace*{-8.8cm}
\mbox{\epsfig{figure=figures/O8ratio_quench_1.2.eps,angle=0,width=0.475\linewidth}\put(29,0)
{\epsfig{figure=figures/O8ratio_poly_1.2.eps,angle=0,width=0.475\linewidth}}}\\
\end{center}
\caption{\label{fig:fit_result_O8}Plots illustrating the relative
quality of the q$\chi$PT and polynomial fits for $\<\op_8\>$. The
18 points for each of the two values of momentum correspond to the
masses from data set 2 in tab.~\ref{tab:ps_mass} and are plotted
as a function of $M_K$.}
\end{figure}

Of the fits which we have tried the best results have been
obtained using the polynomial fits. This is illustrated in
fig.\,\ref{fig:fit_result_O8}, where we display the ratio of the
lattice amplitude and the fitted amplitude for q$\chi$PT and
polynomial fits. This exercise shows that the NLO polynomial
correction are essential to describe the data in the range we have
simulated, whereas the addition of the chiral logarithms (which
have fixed coefficients) spoils the quality of the fits. Since at
sufficiently low masses the NLO chiral logarithms must be present,
we have also tried the \textit{centaur} procedure in which we
smoothly match the polynomial fits (the \textit{horse}) to the NLO
$\chi$PT formula (the \textit{man}) at some kinematical point. The
details of the matching procedure will be described below. The
centaur procedure allows us to estimate the systematic uncertainty
in the extrapolation of our data to the physical point. Of course,
these uncertainties will not be fully under control until
simulations in full QCD are performed at sufficiently small masses
so that the NLO $\chi$PT formula is seen to hold.

We perform the fits on the two subsets of our data corresponding
to the two independent simulations:
\begin{enumerate}
\item[1)] in set 1 masses and energies have values in the interval
$[0.52,1.5]\gev$; \item[2)] in set 2 their values are restricted
to be in the interval $[0.54,1.2]\gev$. This set was generated in
order to increase the number of points at smaller masses.
\end{enumerate}

We start with a discussion of the evaluation of
$_{I=2}\la\pi\pi|\op_{7,8}|K^{0}\ra$ for which we need only
a relatively small number of low energy constants (five, one at LO
and four more at NLO). In addition the chiral expansion starts at
$O(M_{\pi,K}^0)$ (rather then $O(M_{\pi,K}^2)$) and so the fits
are much more stable and an estimate of the matrix elements is
possible to a reasonable accuracy. As explained above, in addition
to using NLO q$\chi$PT formulae, we also fit the data to the
polynomial obtained by removing the chiral logarithms.

The results obtained using the procedures described above are
presented in tab.~\ref{tab:fit_result_O78_PT_f}. $\langle
\op_{7,8}\rangle_{I=2}^{\textrm{\scriptsize{phys}}}$ represents
the value of the matrix element at physical values of the masses,
energies and momenta at NLO in the chiral expansion. By $\langle
\op_{7,8}\rangle_{I=2}^{\gamma}$ we denote the contribution of the
leading order to the physical result. Note that this is the value
of the matrix element in the chiral limit, which is the quantity
determined from previous studies, in which only $K\rightarrow \pi$
matrix elements were calculated. We observe from
tab.~\ref{tab:fit_result_O78_PT_f} (see also
fig.~\ref{fig:fit_result_O8}) that the polynomial form gives by
far the best description of our lattice data for
$_{I=2}\la\pi\pi|\op_{7,8}|K^{0}\ra$, indicating that the data are
not in the kinematic range where the NLO chiral expansion for
these amplitudes are valid.

From our study we see that the use of q$\chi$PT  does not describe
our lattice data and therefore using fits based on q$\chi$PT alone
can lead to unreasonable results. However, in a more general
context, it has been argued that one may combine the behaviour
expected in chiral perturbation theory at low masses with the
observed polynomial dependence on the kinematical parameters in
lattice simulations ~\cite{Kronfeld:2002ab, Becirevic:2002mh}. We
now attempt to perform such a matching.

\begin{table}[t]
{\small\begin{center}\hspace*{-0.2cm}
\begin{tabular}{c|c|ccc|ccc}\hline\hline
fit ansatz & points & $\langle
\op_7\rangle_{I=2}^{\textrm{\scriptsize{phys}}}$ & $\langle
\op_7\rangle_{I=2}^{\gamma}$ & $\frac{\chi}{d.o.f}$ & $\langle
\op_8\rangle_{I=2}^{\textrm{\scriptsize{phys}}}$ & $\langle
\op_8\rangle_{I=2}^{\gamma}$
& $\frac{\chi}{\textrm{d.o.f}}$ \\
\hline
\hline
\multicolumn{8}{c}{Set 1 (340 configurations)}\\
\hline
poly., $M_K,E_\pi< 1.5$ & 18 & 0.130(15) & 0.155(16) &0.11&0.746(91)&0.802(99)& 0.024\\
poly., $M_K,E_\pi< 1.2$ & 13 & 0.135(15) &0.161(17)&0.13&0.750(93)&0.792(98)&0.020 \\
poly., $M_K,E_\pi< 1.0$ & 10 & 0.140(16) &0.161(17)&0.10&0.753(93)&0.790(98)&0.015 \\
\hline
q$\chi$PT, $M_K,E_\pi< 1.5$ & 18 &0.203(21)&0.308(32)&14.8&1.37(17)&2.06(27)&8.95 \\
q$\chi$PT, $M_K,E_\pi< 1.2$ & 13 &0.302(32)&0.492(51)&5.83&1.51(19)&2.32(30)&3.46 \\
q$\chi$PT, $M_K,E_\pi< 1.0$ & 10 &0.276(29)&0.477(50)&4.86&1.37(18)&2.20(29)&2.88 \\
\hline
\hline
\multicolumn{8}{c}{Set 2 (480 configurations)}\\
\hline
poly., $M_K,E_\pi< 1.2$ & 36 & 0.127(11) &  0.143(12) & 0.022&0.734(64) & 0.761(70) & 0.022\\
poly., $M_K,E_\pi< 1.0$ & 24 & 0.128(12)& 0.143(13) & 0.014  &0.732(68)& 0.754(74) & 0.019\\
poly., $M_K,E_\pi< 0.8$ & 16 & 0.132(13)& 0.142(13)  & 0.013 &0.732(77) & 0.753(77) & 0.025\\
\hline
q$\chi$PT, $M_K,E_\pi< 1.2$ & 36 &0.268(26)&0.425(42)&13.4&1.53(16)&2.33(26)&11.0 \\
q$\chi$PT, $M_K,E_\pi< 1.0$ & 24 &0.265(28)&0.442(47)&7.71&1.47(16)&2.34(27)&6.36 \\
q$\chi$PT, $M_K,E_\pi< 0.8$ & 16 &0.224(25)&0.412(47)&6.06&1.22(14)&2.18(26)&5.01 \\
\hline
\hline
\end{tabular}
\end{center}}
\caption{\label{tab:fit_result_O78_PT_f} Results obtained using
various fitting procedures to our data for the matrix elements
$_{I=2}\la\pi\pi|\op_{7}|K^{0}\ra$ and
$_{I=2}\la\pi\pi|\op_{8}|K^{0}\ra$ renormalised in the
NDR-$\msbar$ scheme at $2 \gev$. Energies and masses are given in
GeV while the matrix elements are in $\gev^3$. When fitting using
the q$\chi$PT formula we take $\alpha=0.1$ and $m_0=0.5\gev$.
Notice that we perform uncorrelated fits, which is the reason for
the very small $\chi^2/$d.o.f. for good fits.}
\end{table}

\begin{figure}
\begin{center}
\epsfig{figure=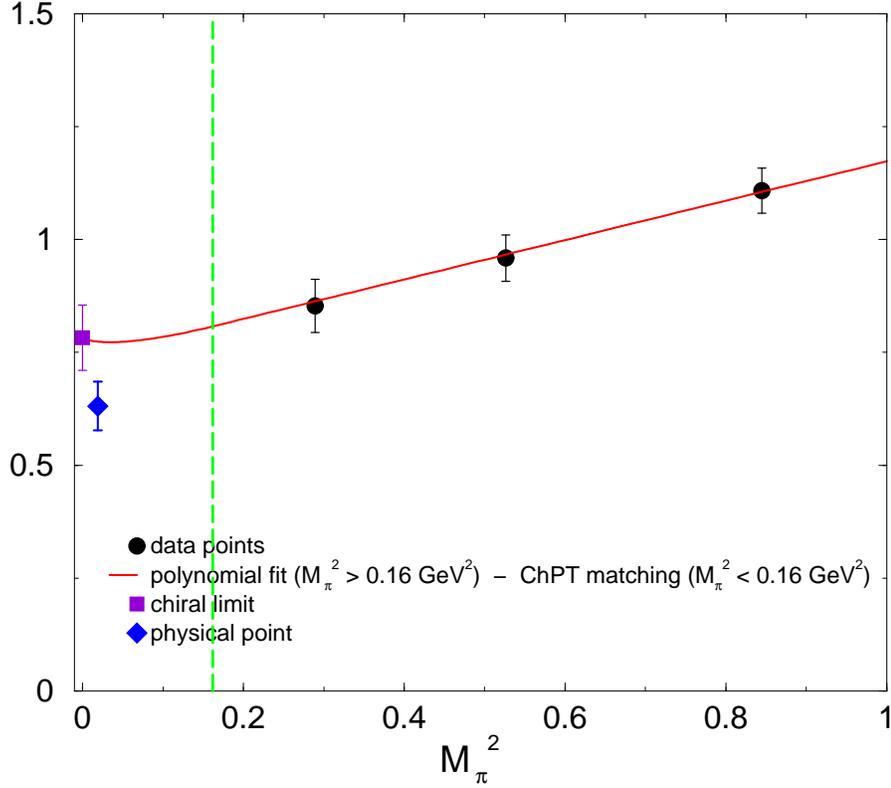,angle=-90,width=0.8\linewidth}
\end{center}
\caption{\label{fig:O8_matching}$\chi$PT-polynomial matching for
$\<\op_8\>_{I=2}$ (in GeV$^3$) at
$(M_{\pi}^\ast,E_{\pi}^\ast,M_{K}^\ast)= (0.4,0.4,0.41)\gev$ (we
plot only data points with $M_{\pi}=E_{\pi}=M_{K}$). For reasons
of numerical stability, it is convenient to use a matching point
in which $M_K^\ast$ is offset from $M_\pi^\ast$ and $M_\pi^\ast$.
The point labelled as \textit{chiral limit} corresponds to
$M_{\pi}=E_{\pi}=M_{K}=0$ and that labelled as \textit{physical
point} corresponds to the physical values of the masses and
energies, and hence does not lie on the SPQR curve.}
\end{figure}

For the most general kinematics the amplitudes $\ampl^{(7,8)}$ are
function of the Lorentz invariants
$p_{\pi^{+}}^{2}=p_{\pi^{0}}^{2}=M_{\pi}^{2}$,
$p_{K}^{2}=M_{K}^{2}$, $p_{K}\cdot p_{\pi^{+}}$, $p_{K}\cdot
p_{\pi^{0}}$ and $p_{\pi^{+}}\cdot p_{\pi^{0}}$, where $p_{K}$,
$p_{\pi^{+}}$ and $p_{\pi^{0}}$ are four-momenta of the kaon and
the two pions. The $I=2$ $|\pi^{+}\pi^{0}\ra$ final state is
symmetric and therefore the dependence on $p_{K}\cdot p_{\pi^{+}}$
and on $p_{K}\cdot p_{\pi^{0}}$ is the same. In the region of
small masses and energies the matrix element is given by $\chi$PT
at NLO, which for $\op_8$, for example, takes the form: \bea
\label{eq:GenKin_KR_chiral_side}
 \ampl^{(8),\chi{\mathrm{PT}}}_{\mathrm{general}}
 &=& A_{(\chi)} \left (1 + [{\mathrm{chiral}}\mbox{ }
  {\mathrm{log}}]_{\mathrm{gen}}(M^{2}_{K},M^{2}_{\pi},p_{\pi^{+}}
  \cdot p_{K},p_{\pi^{0}}\cdot p_{K},p_{\pi^{+}}\cdot p_{\pi^{0}})\right )\nonumber\\
 & & + B_{1(\chi)} M^{2}_{K} + B_{2(\chi)}
   (p_{\pi^{+}}+p_{\pi^{0}})\cdot p_{K}
 + B_{3(\chi)} M^{2}_{\pi}
 + B_{4(\chi)} (p_{\pi^{+}}\cdot p_{\pi^{0}}),
\eea where $A_{(\chi)}$ and the $B_{i(\chi)}$'s are unknown
constants.
On the other hand, in the region accessible to our simulation the
data are well described by a polynomial of the form:
\bea
\label{eq:GenKin_KR_poly_side}
\!\!\!\!\!\!\!\ampl^{(8),\mathrm{poly}}_{\mathrm{general}}
 &=& A_{(p)} + B_{1(p)} M^{2}_{K} + B_{2(p)}
   (p_{\pi^{+}}+p_{\pi^{0}})\cdot p_{K}
 + B_{3(p)} M^{2}_{\pi}
 + B_{4(p)} (p_{\pi^{+}}\cdot p_{\pi^{0}})\,.
\eea
Our procedure is to fit the lattice data to
eq.~(\ref{eq:GenKin_KR_poly_side}), extract $A_{(p)}$ and the
$B_{i(p)}$'s and then to impose the following \textit{smoothness
condition}
\bea
  \left (\ampl^{(8),\chi{\mathrm{PT}}}_{\mathrm{general}}
 \right )_{\mathrm{SPQR\ m.p.}}
 &=& \left (\ampl^{(8),\mathrm{poly}}_{\mathrm{general}}
 \right )_{\mathrm{SPQR\ m.p.}}
 , \nonumber\\
 \left (
 \frac{\partial{\ampl^{(8),\chi{\mathrm{PT}}}_{\mathrm{general}}}}{\partial\, \omega_{i}}
 \right )_{\mathrm{SPQR\ m.p.}}
 &=&
 \left (
 \frac{\partial{\ampl^{(8),\mathrm{poly}}_{\mathrm{general}}}}{\partial\, \omega_{i}}
 \right )_{\mathrm{SPQR\ m.p.}},
\eea where the $\omega_{i},\ i=1$\,--\,4, are the four variables
$M^{2}_{K}, M^{2}_{\pi}, p_{\pi^{+}}\cdot p_{\pi^{0}}$ and
$p_{K}\cdot p_{\pi^{+}}$. The subscript ``SPQR m.p.'' indicates
that the matching point is chosen in the SPQR kinematics and is
thus defined by the values of the masses of the pion and kaon and
the energy of the moving pion
($M_{\pi}^{\ast},M_{K}^{\ast},E_{\pi}^{\ast}$) at which the
matching is performed. This set of conditions determines the
values of $A_{(\chi)}$ and $B_{i(\chi)}$, which can then be used
to compute the physical amplitude using \bea \label{eq:phys_ampl}
 \ampl^{(8),\chi{\mathrm{PT}}}_{\mathrm{phys}} &=&
 A_{(\chi)}\left (1 + [{\mathrm{chiral}}\mbox{ }
{\mathrm{log}}]_{\mathrm{phys}}(m^{2}_{K},m^{2}_{\pi})\right )\nn\\
 && +\left ( B_{1(\chi)} + B_{2(\chi)} +
 \frac{1}{2}B_{4(\chi)} \right ) m_{K}^2
 + \left ( B_{3(\chi)} - B_{4(\chi)} \right ) m_{\pi}^{2} .
\eea Tab.~\ref{tab:O78_KR_12} shows the
results for $_{I=2}\la\pi\pi|\op_{7,8}|K^{0}\ra$ from this
\textit{centaur} procedure, with different choices of matching
points. An example of this procedure is represented graphically in
fig.~\ref{fig:O8_matching}.

\begin{figure}
\begin{center}
\epsfig{figure=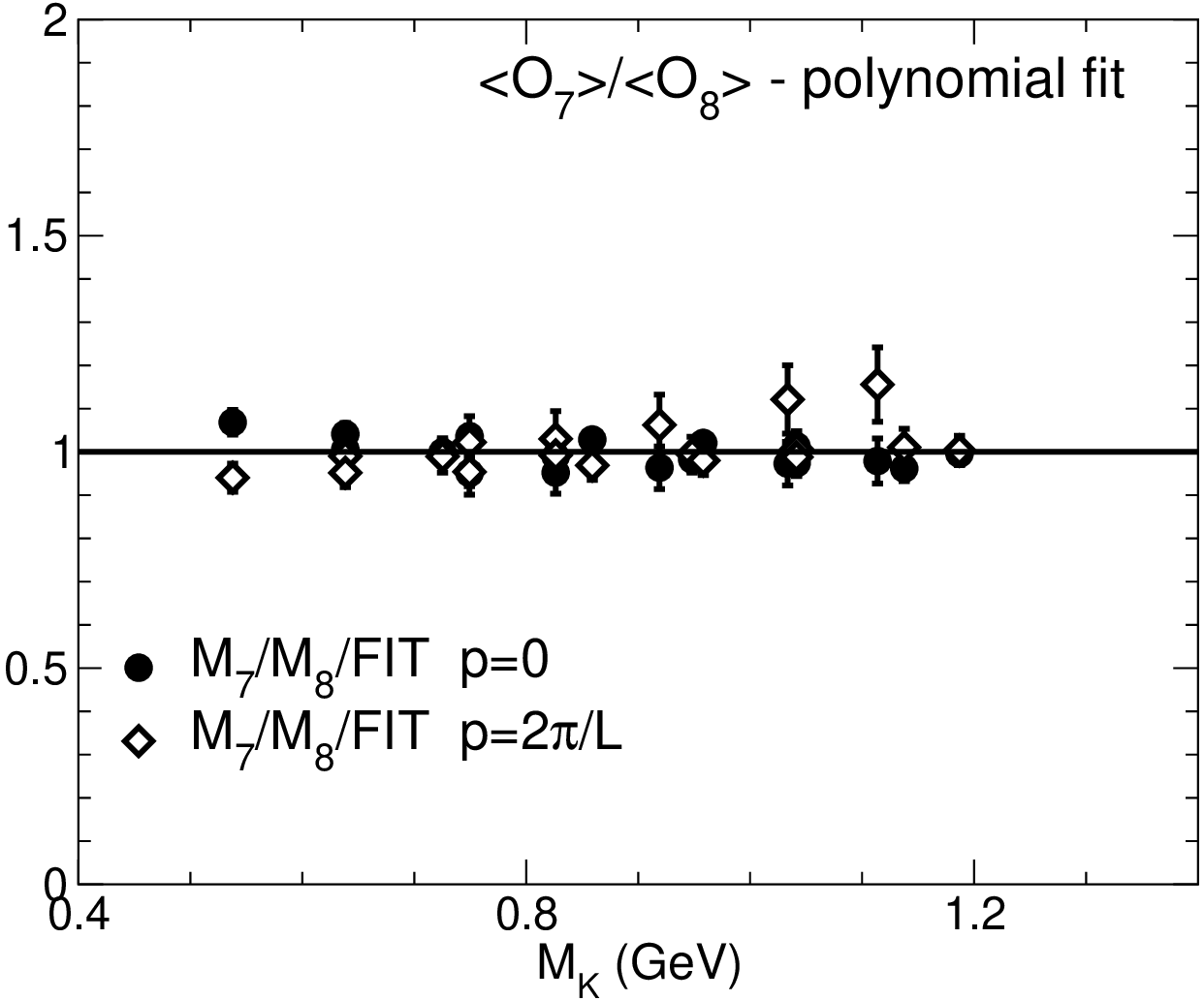,angle=0,width=0.8\linewidth}
\end{center}
\caption{\label{fig:O7sO8}Quality of the polynomial fit for
$\<\op_7\>/\<\op_8\>$.}
\end{figure}

In the ratio ${\cal M}^{(7)}/{\cal M}^{(8)}$ chiral logarithms and
finite-volume corrections cancel since the two operators belong to
the same representation~\cite{Becirevic:2004qd}. This ratio can
also be used to compare lattice results to other theoretical
approaches~\cite{Narison:2000ys,Bijnens:2001ps,Cirigliano:2001qw,Cirigliano:2002jy,Friot:2004ba}.
Results are reported in tab.~\ref{tab:O7sO8}. In contrast to the
case of $\<\op_7\>$ and $\<\op_8\>$ individually, this ratio is
not so well described by the polynomial form coming from $\chi$PT.
This is clear by comparing the values for the $\chi^2/$d.o.f of
tab.~\ref{tab:O7sO8} with those of
tab.~\ref{tab:fit_result_O78_PT_f} and also by comparing
fig.~\ref{fig:O7sO8} with fig.~\ref{fig:fit_result_O8}. We will
discuss these results in sec.~\ref{sec:results}.

Implicit in our work is the assumption that, in the region in
which we have data, the lattice results are reliable even though
the simulations have been performed in the quenched approximation.
When performing the \textit{centaur} matching we join the
polynomial fit of our data with the NLO $\chi$PT formula
corresponding to full QCD. At low masses and energies (where we do
not have data) the behaviour predicted by quenched $\chi$PT is
very different from that in full QCD. However, in spite of the
fact that we have data from a quenched simulation it does not make
sense to extrapolate to the physical point using q$\chi$PT. Our
central value for $_{I=2}\la\pi\pi|\op_{7,8}|K^{0}\ra$ is obtained
by the \textit{centaur} matching, with full chiral logarithms, at
the matching point $(M_{\pi}^*,E_{\pi}^*,M_{K}^*)
=(0.4,0.4,0.5)\gev$. At the same time we also obtain the estimate
of the LECs in full QCD reported in tab.~\ref{tab:lecs}. Other
results reported in these tables are used to estimate systematic
errors, and will be discussed in detail in
sec.~\ref{sec:chiral_error}.

\begin{table}
\hspace*{-0.3cm}
{\small
\begin{tabular}{cc|cc|cc|c}
\hline \hline
 $M_\pi^{*}$& $M_K^{*}$ & $\langle
 \op_7\rangle_{I=2}^{\textrm{\scriptsize{phys}}}$ & $\langle \op_7\rangle_{I=2}^{\gamma}$
& $\langle \op_8\rangle_{I=2}^{\textrm{\scriptsize{phys}}}$ &
$\langle
 \op_8\rangle_{I=2}^{\gamma}$ & $\frac{\<\op_7\rangle_{I=2}^{
\textrm{\scriptsize{phys}}}}{\langle \op_8\rangle_{I=2}^{\textrm{\scriptsize{phys}}}}$\\
\hline\hline
\multicolumn{2}{c|}{poly. fit}
&$0.130(15)\left(^{+8}_{-13}\right)$
 &$0.155(16)\left(^{+8}_{-14}\right)$&
 $0.746(91)\left(^{+49}_{-59}\right)$ &
 $0.802(99)\left(^{+53}_{-64}\right)$ & 0.174\\
\hline
\multicolumn{2}{c|}{poly. fit}
&$0.127(11)\left(^{+7}_{-8}\right)$
 &$0.143(12)\left(^{+8}_{-8}\right)$&
 $0.734(64)\left(^{+44}_{-44}\right)$ &
 $0.761(70)\left(^{+46}_{-46}\right)$& 0.173\\
\hline
\hline
\multicolumn{7}{c}{matching in the SPQR kinematics ($E_\pi^{*}=M_\pi^{*}$) with
full logarithms}\\ \hline
0.3 & 0.5 & $0.127(14)\left(^{+8}_{-13}\right)$
 &$0.154(16)\left(^{+8}_{-14}\right)$&
 $0.728(89)\left(^{+48}_{-58}\right)$ &
 $0.799(99)\left(^{+53}_{-64}\right)$& 0.174\\
0.3 & 0.4 & $0.123(14)\left(^{+7}_{-12}\right)$
 &$0.159(17)\left(^{+9}_{-15}\right)$&
 $0.708(86)\left(^{+47}_{-56}\right)$ &
 $0.821(101)\left(^{+54}_{-66}\right)$ & 0.174\\
0.4 & 0.5 & $0.119(14)\left(^{+7}_{-12}\right)$
 &$0.163(17)\left(^{+9}_{-15}\right)$&
 $0.691(84)\left(^{+46}_{-55}\right)$ &
 $0.843(104)\left(^{+56}_{-67}\right)$& 0.173\\
0.3 & 0.31 & $0.114(13)\left(^{+7}_{-12}\right)$
 &$0.160(17)\left(^{+9}_{-15}\right)$&
 $0.663(81)\left(^{+44}_{-53}\right)$ &
 $0.829(102)\left(^{+55}_{-66}\right)$ & 0.172\\
0.4 & 0.41 & $0.113(13)\left(^{+7}_{-12}\right)$
 &$0.164(17)\left(^{+9}_{-15}\right)$&
 $0.656(80)\left(^{+43}_{-52}\right)$ &
 $0.851(105)\left(^{+56}_{-68}\right)$& 0.172\\
0.5 & 0.51 & $0.113(13)\left(^{+7}_{-12}\right)$
 &$0.170(18)\left(^{+9}_{-16}\right)$&
 $0.660(80)\left(^{+44}_{-53}\right)$ &
 $0.882(109)\left(^{+58}_{-71}\right)$ & 0.171\\
\hline
0.3 & 0.5 & $0.124(11)\left(^{+7}_{-7}\right)$
 &$0.143(12)\left(^{+8}_{-8}\right)$&
 $0.716(62)\left(^{+43}_{-43}\right)$ &
 $0.758(70)\left(^{+45}_{-46}\right)$& 0.173\\
0.3 & 0.4 & $0.121(11)\left(^{+7}_{-7}\right)$
 &$0.147(12)\left(^{+8}_{-8}\right)$&
 $0.698(60)\left(^{+42}_{-42}\right)$ &
 $0.779(72)\left(^{+47}_{-47}\right)$& 0.173\\
0.4 & 0.5 & $0.117(11)\left(^{+7}_{-7}\right)$
 &$0.151(13)\left(^{+8}_{-8}\right)$&
 $0.682(59)\left(^{+41}_{-41}\right)$ &
 $0.800(74)\left(^{+48}_{-48}\right)$& 0.172\\
0.3 & 0.31 & $0.112(10)\left(^{+7}_{-7}\right)$ &
 $0.148(12)\left(^{+8}_{-8}\right)$&$0.655(56)\left(^{+39}_{-39}\right)$&
 $0.786(73)\left(^{+47}_{-47}\right)$ & 0.171\\
0.4 & 0.41 & $0.111(10)\left(^{+6}_{-7}\right)$
 &$0.152(13)\left(^{+8}_{-8}\right)$&
 $0.648(56)\left(^{+39}_{-39}\right)$ &
 $0.807(75)\left(^{+48}_{-49}\right)$ & 0.171\\
0.5 & 0.51 & $0.112(10)\left(^{+7}_{-7}\right)$
 &$0.158(13)\left(^{+9}_{-9}\right)$&
 $0.652(56)\left(^{+39}_{-39}\right)$ &
 $0.836(77)\left(^{+50}_{-50}\right)$ & 0.172\\
\hline
\hline
\multicolumn{7}{c}{matching in the SPQR kinematics ($E_\pi^{*}=M_\pi^{*}$) with
quenched logarithms}\\ \hline
0.3 & 0.5 & $0.135(15)\left(^{+8}_{-14}\right)$
 &$0.184(19)\left(^{+10}_{-17}\right)$&
 $0.772(94)\left(^{+51}_{-62}\right)$ &
 $0.953(117)\left(^{+63}_{-76}\right)$& 0.175\\
0.3 & 0.4 & $0.133(15)\left(^{+8}_{-13}\right)$
 &$0.180(19)\left(^{+10}_{-17}\right)$&
 $0.763(93)\left(^{+51}_{-61}\right)$ &
 $0.936(115)\left(^{+62}_{-75}\right)$ & 0.174\\
0.4 & 0.5 & $0.139(16)\left(^{+8}_{-14}\right)$
 &$0.208(22)\left(^{+11}_{-19}\right)$&
 $0.794(97)\left(^{+53}_{-63}\right)$ &
 $1.07(13)\left(^{+7}_{-9}\right)$ & 0.175\\
0.3 & 0.31 & $0.130(15)\left(^{+8}_{-13}\right)$
 &$0.178(19)\left(^{+10}_{-16}\right)$&
 $0.743(91)\left(^{+49}_{-59}\right)$ &
 $0.923(114)\left(^{+61}_{-74}\right)$ & 0.175\\
0.4 & 0.41 & $0.135(15)\left(^{+8}_{-14}\right)$
 &$0.203(21)\left(^{+11}_{-19}\right)$&
 $0.771(94)\left(^{+51}_{-61}\right)$ &
 $1.05(13)\left(^{+69}_{-84}\right)$ & 0.175\\
0.5 & 0.51 & $0.153(17)\left(^{+9}_{-15}\right)$
 &$0.246(26)\left(^{+13}_{-23}\right)$&
 $0.866(106)\left(^{+57}_{-69}\right)$ &
 $1.27(16)\left(^{+84}_{-102}\right)$ & 0.176\\
\hline
0.3 & 0.5 & $0.132(12)\left(^{+8}_{-8}\right)$
 &$0.170(14)\left(^{+9}_{-9}\right)$&
 $0.758(66)\left(^{+45}_{-46}\right)$ &
 $0.905(84)\left(^{+54}_{-55}\right)$& 0.174\\
0.3 & 0.4 & $0.130(12)\left(^{+8}_{-8}\right)$
 &$0.167(14)\left(^{+9}_{-9}\right)$&
 $0.750(65)\left(^{+45}_{-45}\right)$ &
 $0.888(82)\left(^{+53}_{-54}\right)$& 0.173\\
0.4 & 0.5 & $0.136(12)\left(^{+8}_{-8}\right)$
 &$0.192(16)\left(^{+10}_{-11}\right)$&
 $0.779(68)\left(^{+47}_{-47}\right)$ &
 $1.02(9)\left(^{+6}_{-6}\right)$& 0.175\\
0.3 & 0.31 & $0.127(11)\left(^{+7}_{-7}\right)$
 &$0.165(14)\left(^{+9}_{-9}\right)$&
 $0.731(63)\left(^{+44}_{-44}\right)$ &
 $0.876(81)\left(^{+53}_{-53}\right)$ & 0.174\\
0.4 & 0.41 & $0.132(12)\left(^{+8}_{-8}\right)$
 &$0.188(16)\left(^{+10}_{-10}\right)$&
 $0.757(66)\left(^{+45}_{-46}\right)$ &
 $0.995(92)\left(^{+60}_{-60}\right)$ & 0.174\\
0.5 & 0.51 & $0.149(13)\left(^{+9}_{-9}\right)$
 &$0.227(19)\left(^{+12}_{-13}\right)$&
 $0.848(74)\left(^{+51}_{-51}\right)$ &
 $1.21(11)\left(^{+72}_{-73}\right)$ & 0.176\\
\hline
\hline
\end{tabular}}
\caption{\label{tab:O78_KR_12}Results for $_{I=2}\la\pi\pi|\op_{7,8}|K^{0}\ra$ via
$\chi$PT-polynomial matching performed with full and quenched
chiral logarithms from Set 1 (340 configurations), first block of results, and Set 2 (480 configurations),
second block.
The matrix elements are renormalised in the NDR-$\msbar$
scheme at $2 \gev$. Energies and masses are in GeV
while the matrix elements are in $\gev^3$. In the q$\chi$PT formula we
use $\alpha=0.1$ and $m_0=0.5\gev$.}
\end{table}
%
%
\begin{table}
{\small\begin{center}\hspace*{-0.2cm}
\begin{tabular}{c|c|cc}\hline\hline
data sets & points & $\langle \op_7\rangle_{I=2}^
{\textrm{\scriptsize{phys}}}/\langle
\op_8\rangle_{I=2}^{\textrm{\scriptsize{phys}}}$ &
$\frac{\chi}{d.o.f}$ \\
\hline
\hline
Set 1, $M_K,E_\pi< 1.5$ & 18 & $0.157(12)\left(^{+2}_{-5}\right)$ & 3.11\\
Set 1, $M_K,E_\pi< 1.0$ & 10 & $0.171(11)\left(^{+2}_{-3}\right)$ & 3.02\\
\hline
\hline
Set 2, $M_K,E_\pi< 1.2$ & 36 & $0.153(10)\left(^{+3}_{-2}\right)$ & 1.33\\
Set 2, $M_K,E_\pi< 0.8$ & 16 & $0.163(10)\left(^{+2}_{-2}\right)$ & 1.14\\
\hline
\hline
\end{tabular}
\end{center}}
\caption{\label{tab:O7sO8} Results of the polynomial fit for the ratio
${\cal M}^{(7)}/{\cal M}^{(8)}$ where the operators are renormalised in the
NDR-$\msbar$ scheme at $2 \gev$.}
\end{table}
\begin{table}
{\small\begin{center}\hspace*{-0.2cm}
\begin{tabular}{c|cc|cc}\hline\hline
LECs & Set 1 $\op_7$ & Set 2 $\op_7$ & Set 1 $\op_8$ & Set 2 $\op_8$ \\
\hline
\hline
$\gamma/f^3 [\rm{GeV}^3]$& 0.0705(75)&0.0653(54) &0.365(45) &0.347(32) \\
$(4 \pi f)^2 (\delta_1+\delta_2)/\gamma$& -0.42(90)& -1.08(59)& -1.20(64)&-1.64(53)\\
$(4 \pi f)^2 (\frac{\delta_2}{2}-\frac{\delta_1}{2}+\delta_3)/\gamma$&
-0.46(33)&-0.36(31)&-1.27(20) &-1.35(19)\\
$(4 \pi f)^2 (\delta_4+\delta_5)'/\gamma$& 0.12(43)&-0.06(24) &0.66(26) & 0.44(21)\\
$(4 \pi f)^2 \delta_6'/\gamma$&-0.59(18) &-0.39(13) &-0.92(13) & -0.75(11)\\
\hline \hline
\end{tabular}
\end{center}}
\caption{\label{tab:lecs} Results for the LECs of
$_{I=2}\la\pi\pi|\op_{7,8}|K^{0}\ra$  (where operators are
renormalised in the NDR-$\msbar$ scheme at $2 \gev$) from the {\it
centaur} matching with $(M_{\pi}^*,E_{\pi}^*,M_{K}^*)
=(0.4,0.4,0.5)\gev$. Only statistical errors are displaied.  The
values for $f$ are $0.139(5)\gev$ from set 1 and $0.135(4)\gev$
from set 2. For the definitions of $(\delta_4+\delta_5)'$ and
$\delta_6'$ see eq.~(\ref{eq:redef}).}
\end{table}

\begin{figure}
\begin{center}
\epsfig{file=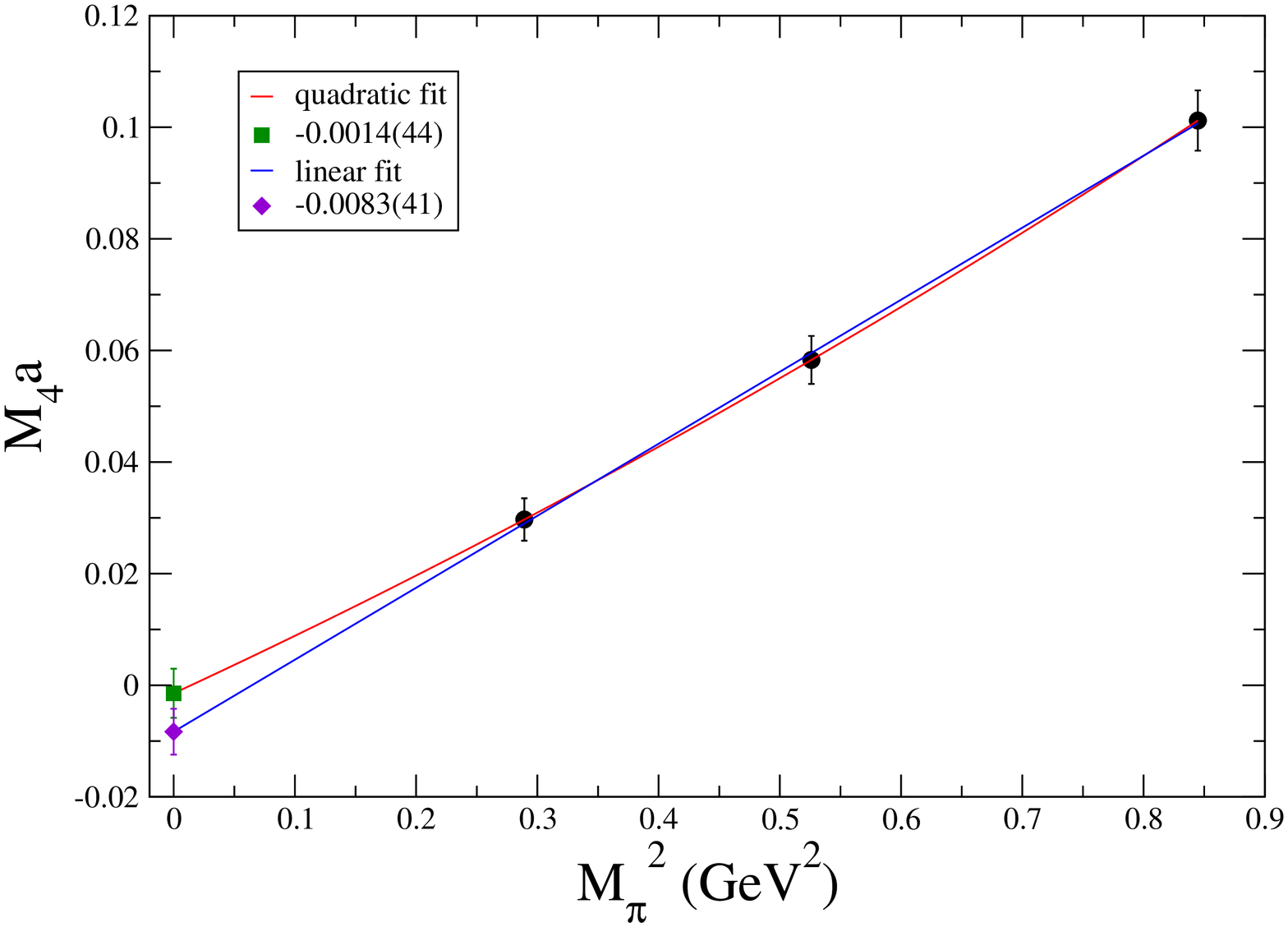,height=12.5cm,width=9.5cm,angle=-90}
\caption{\label{fig:com_o4} {\sl Chiral behaviour of the bare
amplitude (in lattice units) $\bar {\cal M}^{(4)}_{\rm SPQR} (\vec
0)$ for pions with degenerate quarks (lines 1a, 1b, 1c of Tab.3
for dataset 2) versus $M_\pi^2$. The numbers in the box correspond
to the extrapolated values in the chiral limit.}}
\end{center}
\end{figure}

We now turn to the matrix elements of the operator $\op_4$. There
is one LEC at leading order in $\chi$PT and six more at NLO, so
that the fits to the NLO chiral expansion have seven parameters.
Again we find that our data is not well represented by quenched
chiral perturbation theory at NLO. If instead we remove the chiral
logarithms from the fit, we do obtain fits with reasonable values
of $\chi^2$. The coefficients of the polynomial terms at NLO
(which would correspond to the NLO LECs if the chiral logarithms
had been included) are very poorly determined however, making any
extrapolation to the chiral limit impossible. On the other hand,
the coefficient of the leading term is found to be stable, and is
significantly lower (by about 40\%) than the value obtained
keeping only LO chiral perturbation theory. Thus the contribution
of the NLO terms is clearly visible even if, due to the high
number of free parameters in the fit, it is not possible to
evaluate the corresponding LECs with sufficient precision. We
therefore do not discuss further the evaluation of the matrix
elements of $\op_4$ beyond showing one indicative plot. In
fig.\,\ref{fig:com_o4} we plot the values of the bare matrix
element $\bar{\cal M}^{(4)}_{\rm SPQR} (\vec 0)$ from dataset 2 at
the three values of the pion mass with degenerate quarks. Also
shown are linear and quadratic fits to these three points, from
which it can be seen that the extrapolated value in the chiral
limit is close to zero as expected from chiral perturbation theory
(indeed the quadratic fit gives a value consistent with zero).
However, until we have lattice data in the chiral regime, plots
such as that in fig.\,\ref{fig:com_o4} are at best qualitative.

\section{Systematic uncertainties}
\label{sec:systematics}
We now discuss the main systematic errors which affect the values for
the renormalized matrix elements of the electroweak penguins obtained in
this work.

\subsection{Uncertainties from the chiral extrapolation}
\label{sec:chiral_error}
Our results indicate that we are not in a regime where q$\chi$PT
is valid. Completely reliable results can only be obtained by
performing full QCD simulations in a regime where $\chi$PT
describes the data and can therefore be used for the extrapolation
to the physical point. In the meantime we use the centaur
procedure of matching the polynomials which describe our data to
the NLO chiral expansion for
$\la\pi^{+}\pi^{0}|\bar{\op}^{3/2}_{7,8}|K^{+}\ra$. This procedure
is valid under the assumption that there is an overlap region in
which both descriptions hold. We estimate the corresponding
systematic error by varying the matching point below our data down
to the chiral limit (the results are insensitive to varying the
cut-off on the masses and energies of our data points).

\subsection{Uncertainties from operator matching}
\label{sec:op_match_sys} The main source of systematic error in
the evaluation of the renormalization constants is the residual
dependence on the renormalization scale $\mu$ of ${\cal
Z}_+^{RGI}$ and $\hat{\cal Z}^{RGI}$ in eq.~(\ref{eq:appdRGI}). As
explained in sec.\,\ref{subsec:rgbehaviour}, this residual
dependence is probably due to higher order corrections not taken
into account by the NLO perturbative evaluation of the evolution
functions $w_+$ and $\hat w$. Since the scale $\mu=2\gev$ at which
we evaluate the renormalization constants lies inside the range of
momenta at which we compute the renormalization constants, we
estimate the systematic error in the following way: we take the
renormalization constants computed non-perturbatively at a given
scale $\mu_0$ and we run them up or down to the required scale
$\mu$. The systematic error is then estimated from the deviations
of the results obtained by varying $\mu_0$ in the whole range in
which we compute numerically the renormalization constants (i.e.
$\mu_0\in[1.6,2.75]\gev$). In order to reduce this uncertainty we
would need to perform simulations at larger values of $a^{-1}$.
\subsection{Uncertainties from the \boldmath{$\pi{-}\pi$} phase shift and
the finite-volume effects} \label{sec:phase_shift_error}

In this study we have used matrix elements computed with the SPQR
kinematics for which the two-pion system is not at rest. The
Lellouch-L\"uscher (LL) factor relating the finite-volume matrix
elements to physical amplitudes is valid in the centre-of-mass
frame, and the corresponding formula for the moving frame is
currently unknown. This introduces a systematic error and
underlines the importance of developing a theory of finite-volume
effects for two-meson states in a moving frame. Using the LL
factor as a guide with phase-shifts estimated using $\chi$PT we
expect these corrections to be of order 10-15\%, but their effect
on the chiral extrapolation is difficult to estimate.

As shown in sec.~\ref{sec:extrme} and \ref{sec:deltaW}, we were
able to eliminate the systematic error coming from the shift
$\Delta W$ of the two-pion energy in a finite volume. After
applying this correction, since we divide the correlator
$C^{(i)}_{+} + C^{(i)}_{0}$ by two non-interacting pion
propagators (see eq. (\ref{eq:ratio})), we obtain the matrix
element multiplied by the cosine of the phase-shift,
$\cos(\delta^{I=2}(W))$. For the data points in which both pions
are at rest, $\delta^{I=2}(W)=0$ and there is no correction. For
those where one of the two pions has non zero momentum (i.e. in a
moving frame), a first attempt to study the relation between the
finite-volume energy shift and the infinite volume phase-shift has
been presented in ref.~\cite{Rummukainen:1995vs} but a thorough
understanding of this problem proves to be very involved and might
not be resolved soon. For this reason we give here only an
approximate estimate of $\cos(\delta^{I=2}(W))$ based on the
computation of the phase shifts obtained with $N_f=2$ dynamical
simulations in the laboratory
frame~\cite{Yamazaki:2004qb}\footnote{In view of the fact that the
phase shifts for pions in the center of mass frame are very
similar in the dynamical and in the quenched case at the same
lattice spacing and on the same
volume~\cite{Yamazaki:2004qb,Aoki:2002ny}, we assume that this is
the case also in the laboratory frame, for which only dynamical
studies exists.}. For all of the pion masses in our simulation
$\cos(\delta^{I=2}(W))$ turns out to be larger than $0.98$.

\subsection{Quenching effects} \label{sec:quenching_effects}

A reliable estimation of the quenching effects can only be
obtained when the unquenched (or partially quenched) lattice data
are available. The matrix elements presented in this paper have
dimension three, therefore suffer from the uncertainties in the
determination of the lattice spacing, which is one effect of the
quenched approximation. By performing phenomenological studies
using ratios of matrix elements it may be hoped that the
uncertainty is reduced, but it must be remembered that quenching
errors remain and that simulations with dynamical quarks will be
required to eliminate them.

\section{Final results}
\label{sec:results}
Our final results from the two data sets (at the precision of NLO
in the chiral expansion) are
\bea _{I=2}\la
\pi\pi|\op_8|K^0\rangle^{\textrm{\scriptsize{NLO}}}_{\textrm{\scriptsize{phys}}}&=&
0.691(84)\left(^{+46}_{-55}\right)\left(^{+55}_{-35}\right)\gev^3\nn\\
_{I=2}\la
\pi\pi|\op_7|K^0\rangle^{\textrm{\scriptsize{NLO}}}_{\textrm{\scriptsize{phys}}}&=&
0.119(14)\left(^{+7}_{-12}\right)\left(^{+11}_{-6}\right)\gev^3
\label{eq:nlo_bare_ewp_result} \eea from set 1 and \bea _{I=2}\la
\pi\pi|\op_8|K^0\rangle^{\textrm{\scriptsize{NLO}}}_{\textrm{\scriptsize{phys}}}&=&
0.682(59)\left(^{+41}_{-41}\right)\left(^{+52}_{-34}\right)\gev^3\nn\\
_{I=2}\la
\pi\pi|\op_7|K^0\rangle^{\textrm{\scriptsize{NLO}}}_{\textrm{\scriptsize{{phys}}}}&=&
0.117(11)\left(^{+7}_{-7}\right)\left(^{+10}_{-6}\right)\gev^3
\eea from set 2,
where the matrix elements are renormalized in the NDR-$\msbar$
scheme at $2\gev$~\footnote{The normalization of states and
definition of operators in the Introduction implies that \[
\label{eq:o78_32_to_I2}
 _{I=2}\la\pi\pi |\op_{7,8}|K^{0}\ra = \sqrt{\frac{2}{3}}
 \la\pi^{+}\pi^{0} |\op_{7,8}|K^{+}\ra = \sqrt{\frac{1}{6}}
 \la\pi^{+}\pi^{0} |\op^{3/2}_{7,8}|K^{+}\ra.\]}. The first error
is statistical, the second is due to the uncertainties in the
non-perturbative matching to the continuum renormalization scheme
and the third is due to the $\chi$PT-polynomial matching. In
estimating the third error, we have taken the central value as
that obtained by matching to full chiral logarithms at
$(M_\pi^\ast,M_K^\ast)=$ (0.4\,GeV,\,0.5\,GeV), the lower limit to
be the value obtained by matching at (0.4\,GeV, 0.41\,GeV) and the
upper limit to be the value obtained from the polynomial fit. With
this choice, the error bar spans the range of values in the top
half of table\,\ref{tab:O78_KR_12}. The results from the two data
sets are consistent and
as our best estimate we take%
\bea _{I=2}\la
\pi\pi|\op_8|K^0\rangle^{\textrm{\scriptsize{NLO}}}_{\textrm{\scriptsize{phys}}}&=&
0.68\,(6)\,(4)\,(5)\gev^3\nn\\
_{I=2}\la
\pi\pi|\op_7|K^0\rangle^{\textrm{\scriptsize{NLO}}}_{\textrm{\scriptsize{phys}}}&=&
0.12\,(1)\,(1)\,(1)\gev^3\,. \label{eq:best}\eea

Taking the ratio of these two matrix elements directly at the
physical point we obtain the values \bea \frac{_{I=2}\la
\pi\pi|\op_7|K^0\rangle^{\textrm{\scriptsize{NLO}}}_{\textrm{\scriptsize{phys}}}}{_{I=2}\la
\pi\pi|\op_8|K^0\rangle^{\textrm{\scriptsize{NLO}}}_{\textrm{\scriptsize{phys}}}}&=&
0.173(17)\left(^{+2}_{-4}\right)\left(^{+1}_{-2}\right)\qquad
\textrm{from set 1}\nn\\
\frac{_{I=2}\la
\pi\pi|\op_7|K^0\rangle^{\textrm{\scriptsize{NLO}}}_{\textrm{\scriptsize{phys}}}}{_{I=2}\la
\pi\pi|\op_8|K^0\rangle^{\textrm{\scriptsize{NLO}}}_{\textrm{\scriptsize{phys}}}}&=&
0.172(14)\left(^{+2}_{-2}\right)\left(^{+1}_{-1}\right)\qquad
\textrm{from set 2}\,. \label{eq:nlo_bare_ratio_result} \eea The
last error is again due to the uncertainty in the
$\chi$PT-polynomial matching. From the last column of
tab.~\ref{tab:O78_KR_12} one can
appreciate how this ratio is independent of the effects of the
chiral logarithms, i.e. from the choice of the  matching point
$(M_{\pi}^*,E_{\pi}^*,M_{K}^*)$ and from the use of $\chi$PT or
q$\chi$PT formulae. The results in
eq.~(\ref{eq:nlo_bare_ratio_result}) can be compared with those
obtained from the polynomial fit of the ratio ${\cal
M}^{(7)}/{\cal M}^{(8)}$ in tab.~\ref{tab:O7sO8} (which has a
relatively poor $\chi^2$). By reducing the range of masses and
energies used in the fit, its quality increases and the values for
the ratio get closer to those reported in
eq.\,(\ref{eq:nlo_bare_ratio_result}).

The contribution of the leading order term to the matrix elements, which
also corresponds to their value in the chiral limit, is
\bea
_{I=2}\la \pi\pi|\op_8|K^0\rangle_{\gamma}&=&
0.843(104)\left(^{+56}_{-67}\right)\left(^{+40}_{-45}\right)\gev^3\nn\\
_{I=2}\la \pi\pi|\op_7|K^0\rangle_{\gamma}&=&
0.163(17)\left(^{+9}_{-15}\right)\left(^{+7}_{-9}\right)\gev^3
\label{eq:lo_bare_ewp_result} \eea from set 1 and \bea _{I=2}\la
\pi\pi|\op_8|K^0\rangle_{\gamma}&=&
0.800(74)\left(^{+48}_{-48}\right)\left(^{+35}_{-40}\right)\gev^3\nn\\
_{I=2}\la \pi\pi|\op_7|K^0\rangle_{\gamma}&=&
0.151(13)\left(^{+8}_{-8}\right)\left(^{+7}_{-8}\right)\gev^3 \eea
from set 2,
where the sources of errors are the same as in eq.
(\ref{eq:nlo_bare_ewp_result}). From eqs.
(\ref{eq:nlo_bare_ewp_result}) and (\ref{eq:lo_bare_ewp_result}),
it is clear that the NLO contribution of the chiral expansion is
significant. It tends to decrease these matrix elements by about
$15-25\%$. In the kinematical range of the lattice data these NLO
contributions are very significant. In fact, had we computed the
matrix elements at LO in $\chi$PT (i.e. by fitting our data points
to a constant) we would have obtained the following results: \bea
_{I=2}\la
\pi\pi|\op_8|K^0\rangle^{\textrm{\scriptsize{LO}}}_{\textrm{\scriptsize{phys}}}&=&
0.951(74) \gev^3\nn\\ _{I=2}\la\pi\pi|\op_7|K^0
\rangle^{\textrm{\scriptsize{LO}}}_{\textrm{\scriptsize{phys}}}&=&
0.082(9)\gev^3 \eea from set 1 and \bea _{I=2}\la
\pi\pi|\op_8|K^0\rangle^{\textrm{\scriptsize{LO}}}_{\textrm{\scriptsize{phys}}}&=&
0.979(45) \gev^3\nn\\ _{I=2}\la\pi\pi|\op_7|K^0
\rangle^{\textrm{\scriptsize{LO}}}_{\textrm{\scriptsize{phys}}}&=&
0.082(9)\gev^3 \eea from set 2.

\bigskip
\begin{table}
\vspace*{0.5cm}
\begin{center}
\begin{tabular} [c] {c|c|c|c}
\hline \hline
 & $\<\op_7\>_{I=2}^\gamma$ [GeV$^3$]& $\<\op_8\>_{I=2}^\gamma$ [GeV$^3$]  & $\<\op_7\>_{I=2}^\gamma/\<\op_8\>_{I=2}^\gamma$ \\
\hline \hline
This work & $0.16(3)$  & $0.82(15)$ & 0.19(3)\\
\hline \hline
Donini \textit{et al.} \cite{Donini:1999nn}& $0.11(4)$ & $0.51(10)$ & 0.22(9) \\
\hline
RBC \cite{Blum:2001xb} & $0.27(3)$ &  $1.1(2)$ & 0.25(5)\\
\hline
CP-PACS \cite{Noaki:2001un}& $0.24(3)$ & $1.0(2)$ & 0.24(6)\\
\hline \hline
Friot  \textit{et al.} \cite{Friot:2004ba}& $0.12(2)$ & $2.00(36)$ &0.057(15) \\
\hline
Cirigliano \textit{et al.} \cite{Cirigliano:2001qw} & $0.16(10)$ & $2.22(67)$ &0.07(5)\\
\hline
Bijnens \textit{et al.} \cite{Bijnens:2001ps} & $0.24(3)$ & $1.2(8)$ & 0.20(14)\\
\hline
Cirigliano \textit{et al.} \cite{Cirigliano:2002jy}& $0.22(5)$ & $1.50(27)$ & 0.15(4)\\
\hline
Narison \cite{Narison:2000ys}& $0.21(5)$ & $1.40(35)$ & 0.15(5)\\
\hline \hline
\end{tabular}
\end{center}
\caption{Numerical results for the matrix elements of $\op_7$ and
$\op_8$ (renormalised in the NDR-$\msbar$ scheme
at $2 \gev$) in
the chiral limit from various methods (lattice, dispersion
relations, large $N_c$+MHA, sum rules).\label{tab:comparison}}
\end{table}

In tab.~\ref{tab:comparison} we compare our results with those
obtained using different methods (including lattice simulations,
dispersion relations, large $N_c$+MHA and QCD sum rules). None of
the lattice estimates includes the quenching error. The three
previous lattice
computations~\cite{Donini:1999nn,Blum:2001xb,Noaki:2001un} used
$K\rightarrow\pi$ matrix elements and $\chi$PT at leading order to
obtain $K\to\pi\pi$ amplitudes in the chiral limit. The present
work is the first computation with two pions in the final state
which allows the use of $\chi$PT at NLO to extrapolate to the
physical point and to the chiral limit. All of the lattice
computations have been performed at fixed value of the lattice
spacing, but with different actions, different methods of
computing the renormalization factors and different method of
performing the chiral extrapolation, making it very difficult to
compare the systematic uncertainties and to understand the source
of the discrepancy among different results. In the next generation
of computations it will be important to perform simulations at
several values of the lattice spacing in order to study the
approach to the continuum limit, as well as going to lighter
values of the quark masses (the relatively high-values of masses
used in current and previous simulations is one of the major
limitations).

The range of values for $\<\op_8\>_{I=2}^\gamma$ obtained using
lattice simulations is lower than that obtained using other
techniques (although they do overlap within errors). For
$\<\op_7\>_{I=2}^\gamma$ there are discrepancies between the
different lattice simulations and similar discrepancies using
other techniques. Finally, one of the main conclusions of the
present study is that the NLO contribution of $\chi$PT at the
physical point decreases the value of the matrix elements by a
$15-25\%$. This is a physical effect which will affect all the
estimates presented in tab.~\ref{tab:comparison}.
\section{Conclusions}\label{sec:concs}
It can be argued that $K\to\pi\pi$ decay amplitudes represent the
\textit{Holy Grail} of lattice simulations, since so many
theoretical and technical difficulties must be overcome before
they can be calculated reliably. The aim of this study was to get
a better understanding of the practicability of computing
correlation functions with two-pion final states. This is
necessary for extending calculations beyond leading order in
chiral perturbation theory. Specifically in this paper we have
presented the first attempt to calculate the matrix elements for
$K\to\pi\pi$ decays with an $I=2$ final state and to determine all
the low energy constants which appear at next-to-leading order in
the chiral expansion. We were able to carry out all the necessary
steps but with varying degrees of precision and success. Our best
estimates for the matrix elements of the electroweak penguin
operators are presented in sec.\,\ref{sec:results} (see, for
example, eq.\,(\ref{eq:best})). In addition to the quoted errors
should be added those due to the use of the quenched approximation
and finite-volume corrections. Although we could also determine
the matrix elements of the operator ${\cal O}_4$ at the masses
which we used in the simulation, we were unable to determine the
seven low-energy constants with sufficient precision to perform
the chiral extrapolation. For all three operators we conclude that
the NLO terms are significant.

The positive aspects of our exploratory study include:
\begin{enumerate}
\item For the $I=2$ channel, the finite-volume energy shift can be
determined from the linear behaviour of the ratio $R_{4\pi}$
defined in
eq.\,(\ref{eq:ratio4pidef}).%
\item The $K\to\pi\pi$ matrix elements can be determined with good
precision at the values of the masses used in the simulation.%
\item The non-perturbative renormalization of the operators can be
performed successfully.
\end{enumerate}

In addition to removing the quenched approximation by performing
simulations with dynamical quarks, we conclude that two major
improvements are needed before reliable and precise results can be
obtained using the techniques of this paper:
\begin{enumerate}
\item Simulations have to be performed in a region of light-quark
masses where $\chi$PT is valid at NLO for $K\to\pi\pi$ matrix
elements and which is sufficiently large that the low energy
constants can be determined precisely enough for the extrapolation
to the physical point to be possible. We have seen that this is
not the case for the masses used in this simulation and for this
reason we had to use the centaur procedure for the chiral
extrapolation. This needs to be eliminated.%
\item A theory of finite-volume corrections needs to be developed
for two-pion states with a non-zero (total) momentum. For the
spectrum this issue was studied in
ref.\,\cite{Rummukainen:1995vs}. For alternative procedures to
determine the low-energy constants at next-to-leading order, which
exploit more the freedom to vary the quark masses in lattice
simulations rather than introducing a non-zero momentum for the
$\pi\pi$ system see ref.\,\cite{Laiho:2003uy}.
\end{enumerate}
Some minor improvements to our study can be made relatively
easily. For example, the factor of $\cos(\delta)$ due to the
two-pion sink (and the corresponding finite-volume effects) can be
eliminated by computing four-pion correlation functions. This
would also provide a consistency check on the value of the
finite-volume energy shift.

\setlength{\parindent}{0mm}
\begin{Large}
\begin{flushleft}
{\bf Acknowledgements}
\end{flushleft}
\end{Large}

We thank Damir Becirevic, Jonathan Flynn,  Massimo Testa and
Giovanni Villadoro for useful discussions, and Federico Rapuano
for his participation in this work at the early stage. We are
grateful to Alessandro Lonardo, Davide Rossetti and Piero Vicini
for their constant technical support concerning the APEmille
machines.  C.-J.D.L. thanks Universit\'{a} di Roma ``La
Sapienza'', and M.P. thanks the University of Southampton for
hospitality during the progress of this project. CJDL, GM, MP and
CTS thank Maarten Golterman and Santi Peris for hospitality during
the Benasque workshop on \textit{Matching Light Quarks to Hadrons}
where this work was completed.

This work was supported by DOE (USA) grants DE-FG02-00ER41132,
DE-FG03-\-97ER41014 and DE-FG03\--\-96ER40956, European Union
grant HTRN-CT-2000-00145, MCyT Plan Nacional I+D+I (Spain) under
Grant BFM2002-00568, and PPARC (UK) grants PPA/G/S/1998/00529 and
PPA/G/O/2002/00468.
\begin{appendix}

\section{Mass Dependence of $K\to\pi$ Matrix Elements.}

In this appendix we consider the chiral behaviour of a simpler
quantity than the $K\to\pi\pi$ matrix elements described in the
body of the paper, viz. the $K\to\pi$ matrix elements from the
quenched simulation of ref.~\cite{Noaki:2001un}. For illustration
we consider the matrix element of the operator $\op^{3/2}_{8}$ and
show that the uncertainties due to the extrapolation in the quark
mass are also significant in this case.

\begin{figure}
\begin{center}
\epsfig{file=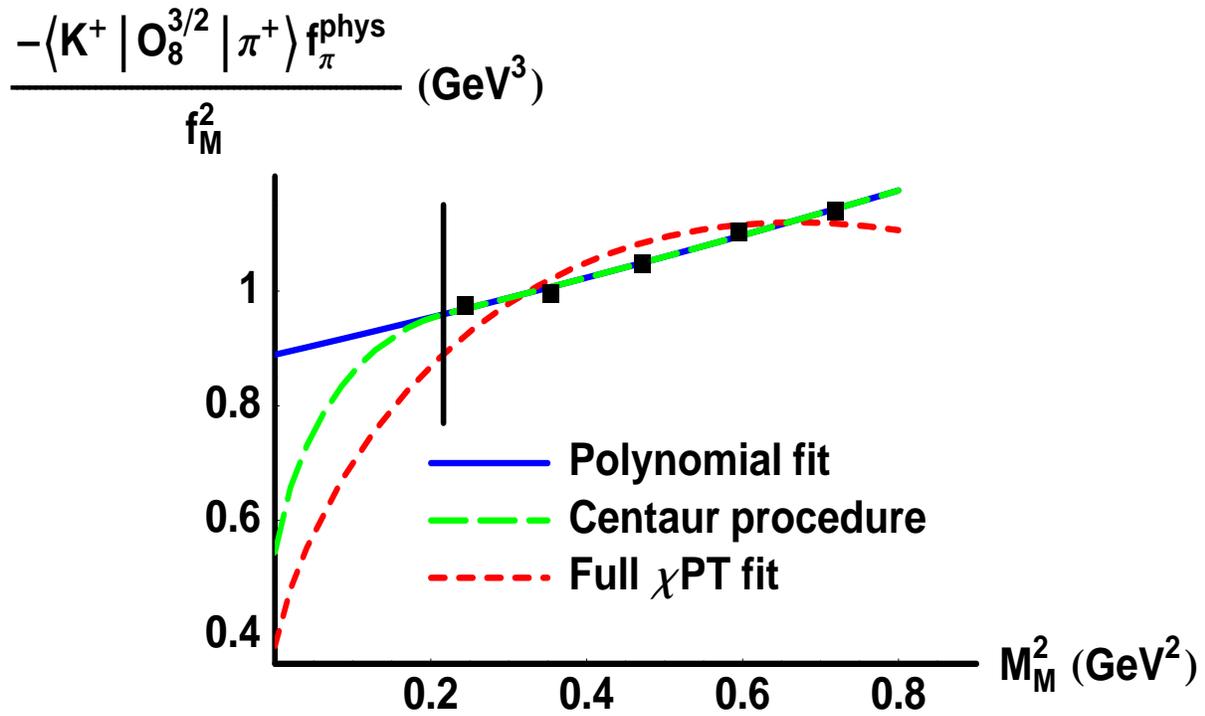,height=12cm,width=16cm}
\caption{$K\to\pi$ matrix element using data from
ref.~\cite{Noaki:2001un}. The $\chi$PT fit corresponds to full
QCD. The vertical line indicates the value of $M_{M}=465$ MeV
where the $\chi$PT-polynomial matching is performed.
\label{fig:CP_PACS_data}}
\end{center}
\end{figure}

\begin{figure}
\begin{center}
\epsfig{file=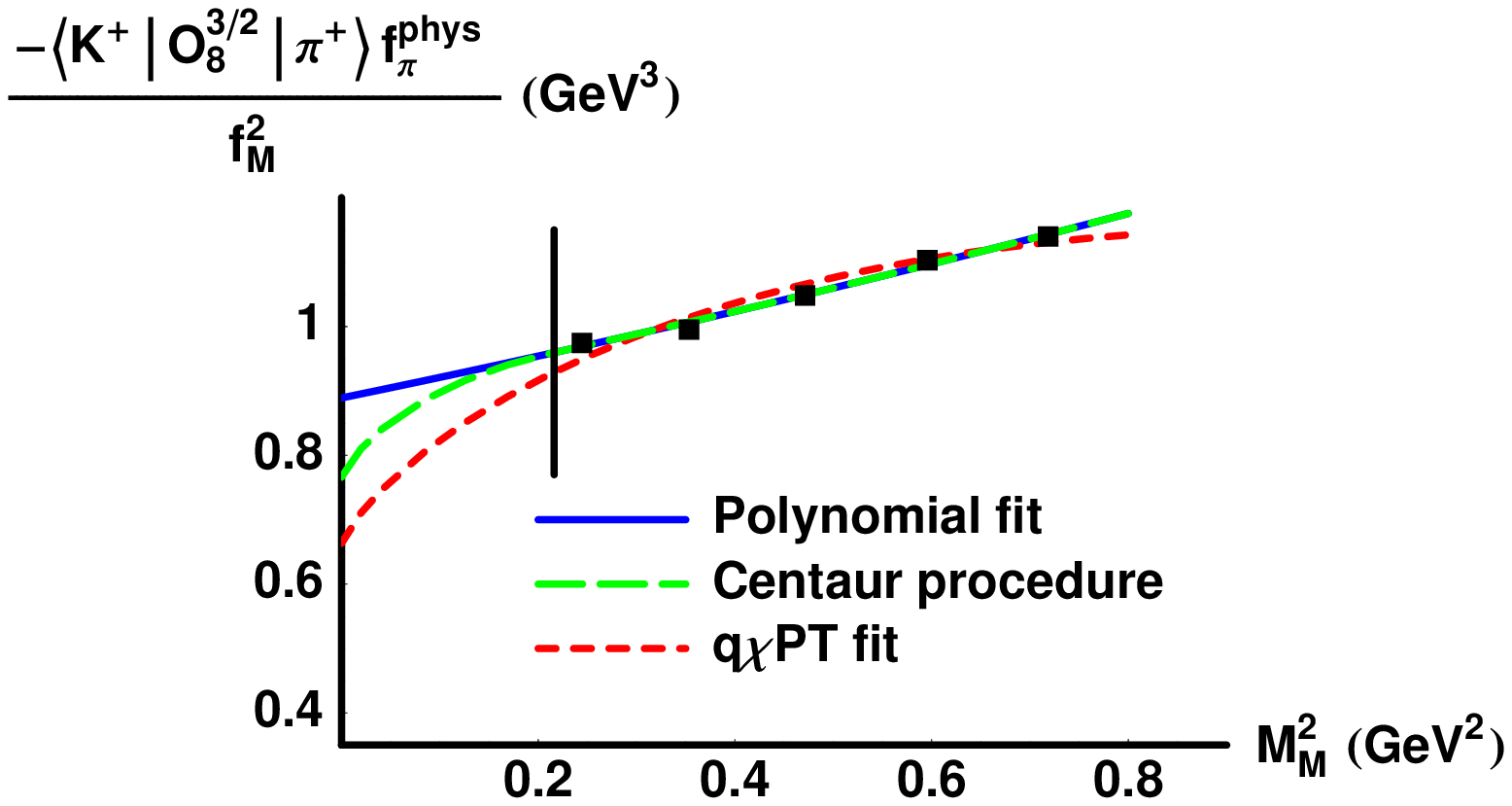,height=12cm,width=16cm}
\caption{$K\to\pi$ matrix element using data from
ref.~\cite{Noaki:2001un}. The $\chi$PT fit corresponds to quenched
QCD. The vertical line indicates the value of $M_{M}=465$ MeV
where the $\chi$PT-polynomial matching is performed.
\label{fig:CP_PACS_DI32_data}}
\end{center}
\end{figure}

In figs.~\ref{fig:CP_PACS_data} and \ref{fig:CP_PACS_DI32_data},
we present the results obtained using different ans\"{a}tze for
the chiral fits to the quenched data for the ratio
$\la\pi^{+}|\op^{3/2}_{8}|K^{+}\ra f_{\pi}/f_{M}^{2}$ (taken from
ref.\cite{Noaki:2001un}). The matrix element
$\la\pi^{+}|\op^{3/2}_{8}|K^{+}\ra$ is determined for
$M_{K}=M_{\pi}\equiv M_{M}$.  $f_{\pi}\approx 132$ MeV is the
physical pion decay constant, and $f_{M}$ is the decay constant of
the pseudoscalar meson with the mass $M_{M}$. In order to estimate
the physical result we perform the following procedures:
\begin{enumerate}
 \item We fit the lattice data to a quadratic function of the form
       $A^{(p)}_{0} + A^{(p)}_{2}M^{2}_{M} +
       A^{(p)}_{4}M^{4}_{M}$. The fit is shown in
       figs~\ref{fig:CP_PACS_data} and \ref{fig:CP_PACS_DI32_data}.
 \item We also fit the data to the prediction of $\chi$PT of the form
$A^{(\chi)}_{0}(1 + [{\mathrm{chiral}}\mbox{ }{\mathrm{log}}])+
A^{(\chi)}_{2}M^{2}_{M}$.
    The chiral logarithms in full QCD
    for $\la\pi^{+}|\op^{3/2}_{8}|K^{+}\ra$ can be obtained from ref.
    \cite{Cirigliano:1999pv}. In the quenched
    theory, the corresponding chiral logarithms have been
    calculated in ref.~\cite{giovanni}. These fits in full and quenched QCD
    are also shown in figs.~\ref{fig:CP_PACS_data} and
    \ref{fig:CP_PACS_DI32_data} respectively.
\item Finally we carry out the centaur procedure of matching the
polynomial fits to the lattice data (which again are good
representations of the data) to chiral perturbation theory at
lower masses. In the same way as for the $K\to\pi\pi$ matrix
elements discussed in sec.\,\ref{sec:analysis}, we assume that
there is a region where both the polynomial and $\chi$PT behaviour
are valid so that by matching the two forms at some scale (which,
in figs.~\ref{fig:CP_PACS_data} and \ref{fig:CP_PACS_DI32_data} is
chosen to be 465\,MeV) we can obtain the coefficients
$A^{(\chi)}_{i}$ of the chiral expansion from the fitted
coefficients $A^{(p)}_{i}$. We perform the matching by equating
the value of $\la\pi^{+}|\op^{(3/2)}_{8}|K^{+}\ra
f_{\pi}/f_{M}^{2}$ and its first derivative with respect to
$M^{2}_{M}$ at the matching scale.
\end{enumerate}
These figures show that the choice of the ansatz for performing
chiral extrapolation can indeed result in large systematic errors,
because the lattice simulation is carried out at relatively heavy
quark masses. See for example fig.~\ref{fig:CP_PACS_data}, where
the matrix element in the chiral limit using the polynomial fit
and extrapolation is about twice that obtained with the centaur
procedure.


\end{appendix}

\newpage
\addcontentsline{toc}{chapter}{Bibliography}
\bibliographystyle{prsty}
\bibliography{refs}

\end{document}